\documentclass[a4paper,onecolumn,accepted=2021-11-15]{quantumarticle}
\pdfoutput=1
\usepackage[utf8]{inputenc}
\usepackage{xspace}			   
\usepackage[english]{babel}
\usepackage[T1]{fontenc}
\usepackage{setspace}
\usepackage{ragged2e}
\usepackage[all]{xy}
\usepackage{graphicx} 
\usepackage{color}
\usepackage[svgnames]{xcolor}

\usepackage{pdfpages}
\usepackage{amsmath, mathtools}
\usepackage{amssymb,amsfonts,dsfont}
\usepackage{mathrsfs}
\usepackage{amsthm}
\usepackage{bm}				
\usepackage{bbm,upgreek}
\usepackage{makecell}
\usepackage{cellspace}
\usepackage[sort&compress,numbers]{natbib}
\usepackage{url}

\newcommand{\bra}[1]{\langle #1 |} 
\newcommand{\ket}[1]{| #1 \rangle } 
\definecolor{cbl}{rgb}{0,0,1}
 
\definecolor{crd}{rgb}{1,0,0}
 
\newcommand{\upd}{\mathrm{d}}
\newcommand{\tr}{\mathrm{tr}}
\newcommand{\xb}{\mathbf{x}}

\newcommand{\wb}{\mathbf{w}}
\newcommand{\ie}[0]{\textit{i.e.}}
\newcommand{\eg}[0]{\textit{e.g.}}

\newcommand\e{\mathrm{e}}
\newcommand\re{\Re\mathrm{e}}
\newcommand\im{\Im\mathrm{m}}
\newcommand\Hint{\hat{H}_\mathrm{int}}
\newcommand\Hbath{\hat{H}_\mathrm{bath}}
\newcommand\Hsys{\hat{H}_\mathrm{sys}}
\newcommand\normt{{\mathcal{N}(\wb^{[t]},t)}}
\newcommand\norm{{\mathcal{N}(\wb,t)}}
\newcommand\hA{\hat{A}}
\newcommand\hx{\hat{x}}
\newcommand\hp{\hat{p}}
\newcommand\hH{\hat{H}}
\newcommand\lA{\overleftarrow{A}}
\newcommand\rA{\overrightarrow{A}}
\newcommand\sch{Schr\"{o}dinger}

\definecolor{nblue}{rgb}{0.06,0.3,0.73}
\definecolor{nblack}{rgb}{0,0,0}
\definecolor{nred}{rgb}{0.9,0.1,0.1}

\newcommand{\blk}{\color{nblack}}

\title{Non-Markovian wave-function collapse models are Bohmian-like theories in disguise}

\author{Antoine Tilloy}
\email{antoine.tilloy@gmail.com} 
\affiliation{Max-Planck-Institut f\"ur Quantenoptik, Garching, Germany}
\affiliation{Munich Center for Quantum Science and Technology (MCQST), Munich, Germany}

\author{Howard M. Wiseman}

\affiliation{Centre for Quantum Dynamics, Griffith University, Brisbane, Queensland 4111, Australia}
\email{h.wiseman@griffith.edu.au}

\date{}

\begin{document}

\maketitle

\begin{abstract}
Spontaneous collapse models and Bohmian mechanics are two different solutions to the measurement problem plaguing orthodox quantum mechanics. They have, {\em a priori} nothing in common. At a formal level, collapse models add a non-linear noise term to the Schr\"odinger equation, and extract definite measurement outcomes either from the wave function (\eg~mass density ontology) or the noise itself (flash ontology). Bohmian mechanics keeps the Schr\"odinger equation intact but uses the wave function to guide particles (or fields), which comprise the primitive ontology. Collapse models modify the predictions of orthodox quantum mechanics, whilst Bohmian mechanics can be argued to reproduce them. However, it turns out that collapse models and their primitive ontology can be exactly recast as Bohmian theories. More precisely, considering (i) a system described by a non-Markovian collapse model, and (ii) an extended system where a carefully tailored bath is added and described by Bohmian mechanics, the stochastic wave-function of the collapse model is exactly the wave-function of the original system conditioned on the Bohmian hidden variables of the bath.  Further, the noise driving the collapse model is a linear functional of the Bohmian variables. The randomness that seems progressively revealed in the collapse models lies entirely in the initial conditions in the Bohmian-like theory. Our construction of the appropriate bath is not trivial and exploits an old result from the theory of open quantum systems. This reformulation of collapse models as Bohmian theories brings to the fore the question of whether there exists  `unromantic' realist interpretations  of quantum theory that cannot ultimately be rewritten this way, with some guiding law.  It also points to important foundational differences between `true' (Markovian) collapse models and non-Markovian models.
\end{abstract}

\section{Introduction}

Bohmian mechanics \cite{bohm1952I,bohm1952II,durr2009,goldstein2016} and collapse models \cite{ghirardi1986,bassi2003,bassi2013review} are two realist approaches  that straight-forwardly solve the measurement problem of non-relativistic orthodox quantum mechanics (OQM). It is unknown if they resemble what (if anything) is really going on ``behind the scenes'', but they are arguably the simplest examples available of  ``unromantic''  \cite{bell1992} mechanistic accounts of quantum predictions. They are also arguably the only single-world approaches for which one understands reasonably well the full story, from a precise microscopic theory all the way to the statistics of macroscopic measurement results.

At first sight, Bohmian mechanics and collapse models take a very different route to solve the measurement problem. Bohmian mechanics posits that particles have positions $\xb(t)$ defined at all time, which flow along a velocity field $\mathbf{v}_{\ket{\psi}}$ constructed from the wave-function $\ket{\psi}$. The unitary evolution of the wave-function is kept unchanged. Definite measurement outcomes are then obtained from the coarse grained positions of the particles making the measurement apparatus.

In collapse models, on the other hand, one adds a non-linear stochastic modification to the unitary evolution of the wave-function. The modification is taken small enough that microscopic dynamics remain almost unchanged, but large enough that macroscopic superpositions in space are quickly collapsed. Usually, the wave-function in configuration space is supplemented by ``local beables'' \cite{bell1976, allori2015} in physical space (mass density or flashes, corresponding to the collapse events) from which macroscopic measurement results are unambiguously extracted. 

Bohmian mechanics keeps the usual quantum predictions unchanged; collapse models seem to inevitably alter the empirical content. Bohmian mechanics is deterministic, with randomness lying in the initial conditions; collapse models are stochastic, with randomness progressively incorporated by noise. But are these theories actually so different?

Superficial similarities between the two approaches have already been noted in the literature. The first is that both approaches are realist in the most trivial sense: they ultimately describe the dynamics of local beables (or `stuff') from which all predictions can in principle be extracted \cite{allori2008,allori2014}. The second is that the evolution for the wave-function in the Ghirardi-Rimini-Weber (GRW) model, the simplest collapse model, approximates, in an appropriate limit, the dynamics of a Bohmian conditional wave function in a setup where a bath is added \cite{toros2016}.

Our objective is to go beyond these insights and show that (continuous) collapse models can be \emph{exactly} rewritten as Bohmian-style theories. This claim could be understood in a weak sense: collapse models have the same empirical content as Bohmian mechanics applied to a carefully chosen system $+$ bath setup. While true and often overlooked, this result is trivial. Rather, we will show, through a quite involved derivation, that the very objects both theories use are in one-to-one correspondence. Namely, for a given collapse model, one can construct 
\begin{enumerate}
    \item a bath of harmonic oscillators coupled to the system, 
    \item bath hidden variables (linear combinations of positions and momenta of its oscillators), and 
    \item Bohmian-style deterministic dynamics for those hidden variables
\end{enumerate} 
such that the typical dynamics of the wave function of the system conditioned on these bath variables is \emph{exactly} the stochastic wave function dynamics of the collapse model. Further, an appropriate linear function of the Bohmian-stlye hidden variables matches \emph{exactly} one natural choice of local beables for the collapse model. Hence the connection is not only empirical, it is ontological. In the interest of fluidity of expression, in the remainder of this paper, we will take the adjective ``Bohmian'' to encompass the ``Bohmian-style'' theories we consider.

 The real utility of the relation we elucidate is for the case of non-Markovian collapse models. These are models with `gentle' collapse, driven by non-white noises, as opposed to the white noise in Markovian collapse models. However, the mathematical construction of these models is far more complicated than merely substituting colored noise for white noise. 

\begin{figure}
    \centering
    \includegraphics[width=1\textwidth]{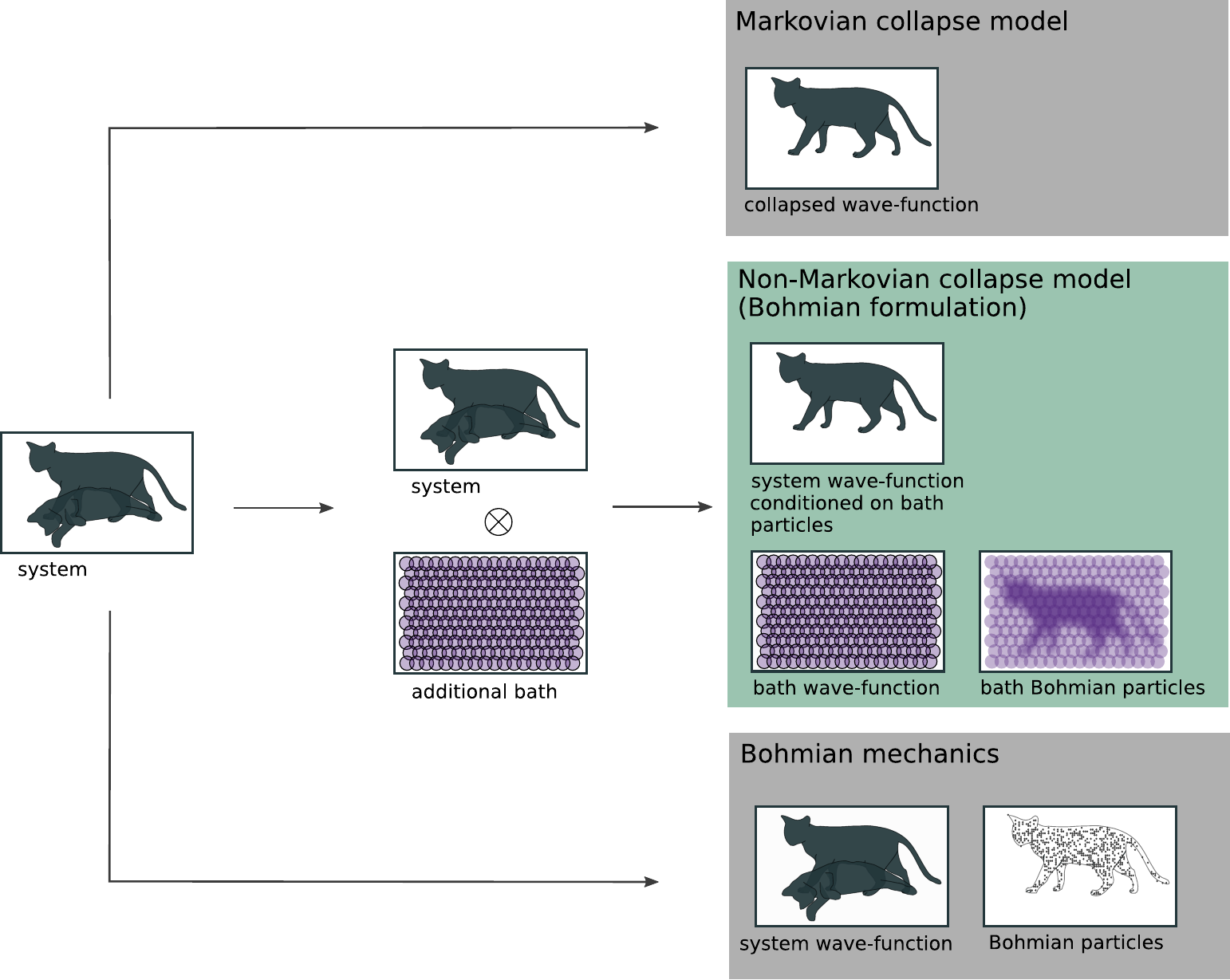}
    \caption{Three realist approaches to solving the quantum measurement problem. According to standard quantum mechanics (left), \sch's equation  implies that the wave-function of a macroscopic system (\eg~a measurement apparatus, or a cat, plus its environment) will evolve into a superposition of macroscopically distinct classical-like states. In Markovian collapse models (top), \sch's equation is modified to cause a collapse into one of these states. In Bohmian mechanics (bottom), the wave-function remains a superposition, but particle positions are added as extra variables, and these will follow trajectories as if only one of those classical-like states pertained. To define consistent non-Markovian collapse models (middle) it is convenient to introduce an additional bath and consider the \sch\ equation for the combined system. We show in this paper that the ``collapsed'' wave-function is most simply defined as the system wave-function conditioned on Bohmian hidden variables for particles in the bath. To define the trajectories of these particles, the uncollapsed joint wave-function of system plus bath is required.}
    \label{fig:interpretations}
\end{figure}

Moreover, the basic ontology in the non-Markovian case is fundamentally different, as the wave-function (even of the whole universe as conventionally conceived) on one time-slice does not carry all information in the past relevant for future evolution. That is because --in our opinion-- the heart of these non-Markovian collapse models is the ancillary  `non-physical' quantum bath, which is conventionally introduced as a mathematical device to ensure \blk consistent quantum evolution for the averaged system state. 
Our mapping to a Bohmian interpretation makes clear that this additional bath actually has a physical role in guiding the collapsing wave-function of the system, and carrying forward relevant information from the past to the future.  The three distinct ontologies of Markovian collapse theories, Bohmian mechanics, and the Bohmian version of non-Markovian collapse theories, are illustrated in Fig.~\ref{fig:interpretations}. 

\newpage 
The essence of the mathematical result we shall derive is already present, albeit in a 
 less focussed  form, in an article \cite{gambetta2003} discussing non-Markovian stochastic Schr\"odinger equations.  
That paper appeared only 1 year after the first paper on collapse models with colored noises \cite{bassi2002}, and prior to their thorough exploration \cite{adler2007,adler2008,bassi2009exactshort,bassi2009exactlong,ferialdi2012exact}, 
which may be why this fundamental mapping has been missed in the collapse literature. Another reason may be that, 
because of the complexity of non-Markovian collapse models,  a correct description of their exact dynamics has often been lacking in the literature (see Sec.~\ref{subsec:NLNMD}). Instead, they  
have been studied mostly perturbatively, 
precluding the elucidation of their exact connection with Bohmian mechanics. 

The remainder of this paper is structured as follows. In Sec.~2 we briefly explain the main result of our paper in relatively non-technical terms. Sec.~3 introduces the conventional mathematical description of collapse models, Markovian and non-Markovian, and Sec.~4 quickly reviews Bohmian mechanics. Thus we are set up to prove the relation between these models in Sec.~5, which ends with a table summarizing the equivalence. We conclude in Sec.~6 with a substantial discussion of generalisations, implications, and other foundational issues. 

\section{A sketch of the equivalence}

Before going into more details, let us sketch in words how the equivalence works and how different quantities are mapped one onto another.  The system could be a system of non-relativistic particles or something more complicated -- the details do not matter. In a non-Markovian collapse model, the system's wave-function  $\ket{\psi_\wb(t)}$ obeys  a quite complicated non-linear equation driven by colored noise  $\wb(t)$. This is a vector because there is one noise term per collapse operator $\hat{A}_k$, and for a macroscopic system there will be many --perhaps continuously infinitely many-- of these. 
 We aim to show how this very general model contains within it an exactly equivalent Bohmian-type model. 

To this end, we first extend the Hilbert space by adding a bath of Harmonic oscillators $\mathscr{H}_\text{tot}~=~\mathscr{H}_\text{sys}~\otimes~\mathscr{H}_\text{bath}$, with a carefully chosen linear coupling.  This produces entanglement between the system and bath, in the joint state $\ket{\Psi(t)}$, 
such that tracing over the bath degrees of freedom is equivalent to a noise averaging of the collapsed state on $\mathscr{H}_\text{sys}$: 
\begin{equation} \label{eq:trace}
    \rho_{\rm sys}(t) = \tr_{\rm bath}[\ket{\Psi(t)}\bra{\Psi(t)}] 
    = \mathds{E}[\ket{\psi_\wb(t)}\bra{\psi_\wb(t)}]
\end{equation}
This condition is necessary to preserve the standard empirical meaning of a quantum state,  avoiding effects such as faster-than-light signalling.

Now the trick is to describe the oscillators in the bath, and them only, as particles with real (in the Bohmian sense) properties, $\xb(t)$. Doing so, we obtain real time deterministic trajectories for them,  guided by $\ket{\Psi(t)}$,  and an associated conditional wave function on the remaining (system) Hilbert space, 
\begin{equation} \label{eq:condB}   
    \ket{\psi_{\xb(t)}(t)} \propto  \bra{\xb(t)}\Psi(t)\rangle \in\mathscr{H}_\text{sys}.
\end{equation}
The evolution of this conditional wave-function is determined by the initial joint wave-function and by the 
the initial condition of the Bohmian particles. 
If the latter are specified only statistically through the Born rule then the evolution of $\ket{\psi_{\xb(t)}(t)}$ can be considered stochastic.  

\begin{figure}
    \centering
    \includegraphics[width=0.82\textwidth]{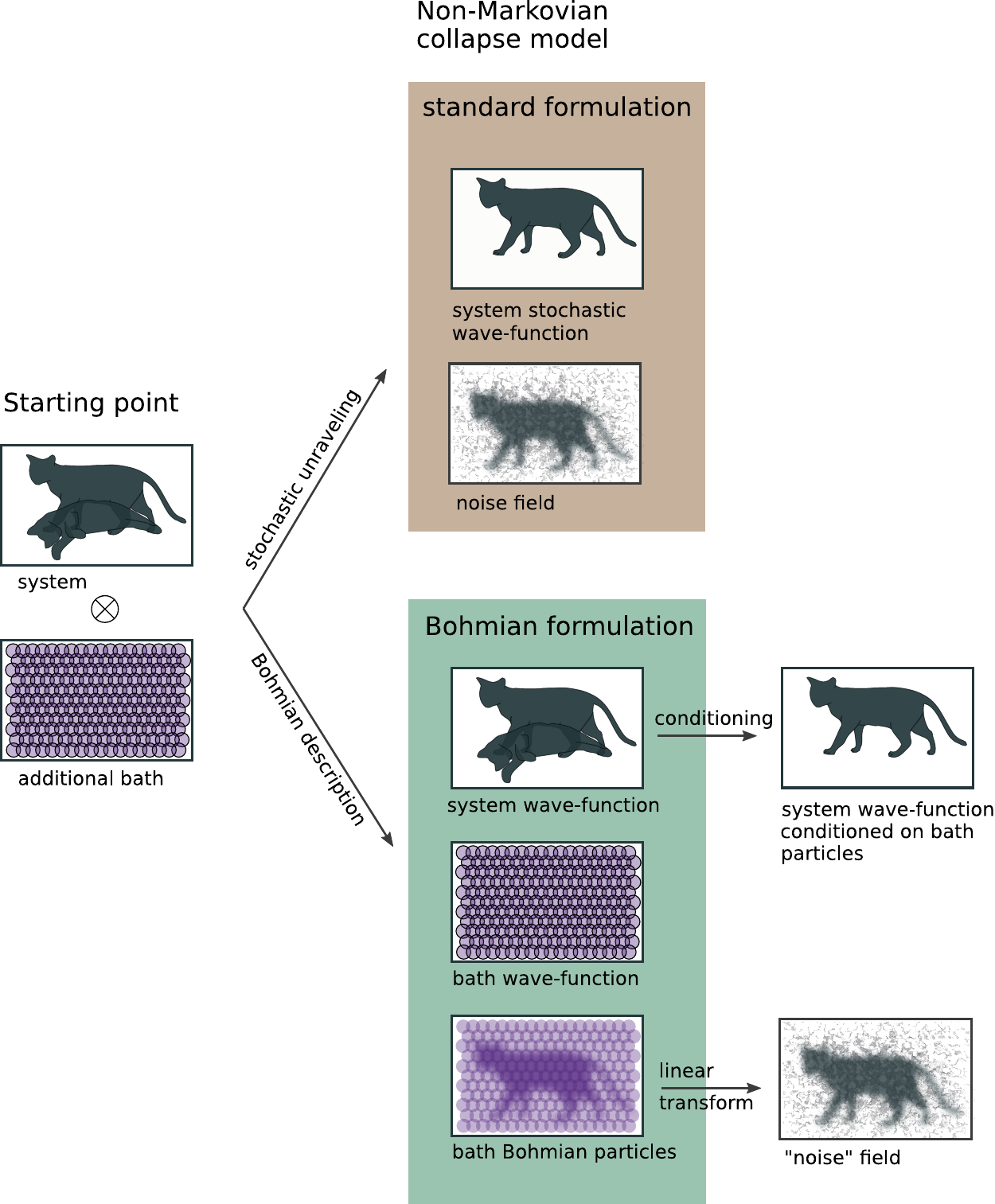}
    \caption{Equivalence of standard (top) and Bohmian (bottom) formulations of non-Markovian ``collapse'' models. As per Fig.~\ref{fig:interpretations}, for such models it is necessary to introduce a bath in addition to the system. In the standard formulation, there is a colored noise process $\wb(t)$, that is a very high-dimensional vector (\eg~3 components per particle) that we visualize as a noise field in space. This noise drives --in a complicated way-- the dynamics of a stochastic system state $\ket{\psi_\wb(t)}$ such that on average it is the same as tracing out the bath, as per  Eq.~(\ref{eq:trace}). In the Bohmian formulation, Bohmian hidden variables $\xb(t)$ are introduced for the bath {\em only}. These are guided deterministically by the deterministically evolving joint wave-function of system and bath. The Bohmian-conditional wave-function $\ket{\psi_{\xb(t)}(t)}$ of the system alone is defined at any time by imagining that the bath were, at that instant, projected into the configuration eigenstate $\ket{\xb(t)}$. As per Eq.~(\ref{eq:condB}), this $\ket{\psi_{\xb(t)}(t)}$ is the {\em same} state as $\ket{\psi_\wb(t)}$, with $\wb(t)$ obtained from $\xb(t)$ by a linear transformation. The stochasticity arises from the Born-random initial conditions for ${\bf x}$.}
    \label{fig:equivalence}
\end{figure}

What we shall prove is that the conditional wave-function $\ket{\psi_{\xb(t)}(t)}$ is exactly the collapse model stochastic wave-function $\ket{\psi_\wb(t)}$; they are the same stochastic process. Further, the colored noise of the collapse model can be  transformed by a linear map   (essentially a Fourier transform) to the Bohmian variables of the oscillators. The randomness of the noise is a progressive unraveling of the randomness of the initial Bohmian variables. The equivalence between the standard and Bohmian description  of non-Markovian collapse models  is summarized schematically in Fig.~\ref{fig:equivalence}.

\section{Collapse models}

\subsection{Markovian models}
A \emph{Markovian} collapse model, driven by white noise is typically given by an It\^o equation of the form:
\begin{equation}\label{eq:sse_markov}
    \upd \ket{\psi(t)} = \left[- i \hH \,\upd t +  \sum_{k=1}^D \left\{\sqrt{\gamma}\,(\hA_k -\langle \hA_k\rangle_t)\,\upd W_k(t) -\frac{\gamma}{2} (\hA_k -\langle \hA_k\rangle_t)^2 \, \upd t \right\}\right] \ket{\psi(t)},
\end{equation}
where $\hH$ is the system Hamiltonian in absence of collapse, $W_k$ are uncorrelated Wiener processes (\ie~$\upd W_k/\upd t$ are independent white noises), $\hA_k$ are the $D$ Hermitian operators  which continuously `collapse' the system,  and $\gamma$ encodes the strength of the collapse term. We keep the system Hilbert  space  generic but, to  aid intuition, it could
be thought of as  that of $N$ particles without spin in a $3$ dimensional space. The operators $\hA_k$ are typically related to position.

As an example, in the Quantum Mechanics with Universal Position Localization (QMUPL) model \cite{diosi1989,bassi2013review} for a single particle, $k\in\{1,2,3\}$ and the $\hA$'s are the 3 coordinate operators $\hat{X}$, $\hat{Y}$, $\hat{Z}$. In the continuous spontaneous localization (CSL) model \cite{pearle1989,ghirardi1990}, the index $k$ is a continuous position in $\mathds{R}^3$ and the $\hA$'s are regularized mass density operators.

The It\^o equation \eqref{eq:sse_markov} is such that  averaging over the noise gives a mixed state $\rho_t = \mathds{E}\left[\,\ket{\psi(t)} \bra{\psi(t)}\,\right]$ that obeys a linear differential equation of the Lindblad form. This is necessary to avoid faster-than-light signalling and a breakdown of the Born rule \cite{gisin1989,gisin1990,polchinski1991,bassi2015}; this is the main condition collapse models have to satisfy and it fixes their form in the Markovian case \cite{bassi2013,wiseman2001}.

\subsection{Non-Markovian linear collapse equation}

Because noise with arbitrarily high frequencies seem unphysical (and potentially make relativistic extensions difficult \cite{myrvold2017,jones2021}) much effort has gone into constructing collapse models with colored noises \cite{bassi2003,adler2007,adler2008}. This is much less trivial than just taking \eqref{eq:sse_markov} and replacing white noise by colored noise, as this would break all its nice properties. Instead, a way to proceed is to introduce a \emph{linear} stochastic Schr\"odinger equation, which may at first sight look quite peculiar:
\begin{equation}\label{eq:sse_linear}
    \frac{\upd}{\upd t} \ket{\phi_\wb(t)} = \left[-i H  + \sqrt{\gamma} \, w_j(t) \hA_j  - 2 \sqrt{\gamma}\,  \hA_j \int_0^t  \upd s \, D_{jk}(t,s) \frac{\delta}{\delta w_k(s)} \right]\ket{\phi_\wb(t)}\, .
\end{equation}
Here, and from now on, there is implicit summation on repeated indices.
The real noise vector $\wb=\{w_k\}_{k=1}^D$ has a Gaussian probability measure $\mathds{Q}$ of zero mean and correlation:
\begin{equation}
    \mathds{E}_\mathds{Q}\left[w_j(t) w_k(s)\right] = D_{jk}(t,s)
\end{equation}
where $D$ is a real positive semi-definite kernel, which depends on $t$ and $s$ only via $t-s$.  The non-Markovianity of collapse models with non-white noises thus is not as benign as one might think: the state is not just driven by colored noise, but rather the time derivative depends on all the past via a functional derivative. Why this special form and more precisely, why the need for this functional derivative term?

As in the Markovian case, for collapse models to make sense, it is important that they yield a legitimate open-system evolution upon noise averaging. This is necessary to maintain the probabilistic interpretation of states (Born rule) and preclude faster-than-light signalling. This is precisely what the functional derivative term achieves. One first notes that equation \eqref{eq:sse_linear} can be formally integrated\footnote{This latter form shows that the solutions of \eqref{eq:sse_linear} can in principle be computed numerically with a diagrammatic expansion or via Monte Carlo \cite{tilloy2017}.} into \cite{diosi2014}:
\begin{equation}\label{eq:sse_integrated}
\ket{\phi_\wb^I(t)} = \mathcal{T}\exp\left\{\sqrt{\gamma}\int_0^t \upd u \, \hA^I_j(u) \left[w_j(u) -2\sqrt{\gamma}\int_0^u \upd v \,D_{jk}(v,u) \hA^I_k(v)\right] \right\}\ket{\phi(0)},
\end{equation}
where $\mathcal{T}$ is the time ordering operator and $\ket{\phi^I_\wb(t)}$ is the state in interaction representation with respect to $\hH$. Then, writing $\rho^I(t) =\mathds{E}_\mathds{Q}\left[ \, \ket{\phi^I_\wb(t)}\bra{\phi^I_\wb(t)} \, \right]$, one finds \cite{diosi2014}, upon Gaussian integration (or Wick's theorem) $\rho^I(t) = \mathcal{M}_t \cdot \rho(0)$, with $\mathcal{M}_t$ a completely positive trace-preserving map:
\begin{equation}\label{eq:master}
   \mathcal{M}_t =\mathcal{T} \exp\bigg\{ \int_0^t\!\!\int_0^t \!\! \upd u\, \upd v \, D_{jk}(u,v) \Big[\lA^k(v) \rA^j(u) - \theta_{u v}\lA^j(u) \lA^k(v)- \theta_{vu} \rA^k(v) \rA^j(u) \Big] \bigg\}\,.
\end{equation}
Here, $\theta_{u v}=\theta(u-v)$ is the Heaviside function and we have used the left-right superoperator notations:
\begin{equation}
\begin{array}{ll}
\overrightarrow{A} \cdot \rho = \hat{A}\rho \; ; &\overleftarrow{A} \cdot \rho = \rho \hat{A}
\end{array}.
\end{equation}
The non-Markovian master equation \eqref{eq:master} is simply an operator rewriting of the Feynman-Vernon influence functional \cite{feynman1963}, usually presented in path integral representation. It means that at the density matrix level, the evolution of the collapse model is that of a system linearly coupled to a bath of harmonic oscillators.

Historically, guessing the form of the linear stochastic differential equation \eqref{eq:sse_linear} was done the other way around \cite{strunz1996,diosi1997}: starting from the legitimate master equation \eqref{eq:master}, `unraveling' it into a stochastic state in integrated form \eqref{eq:sse_integrated}, and deducing its differential form \eqref{eq:sse_linear} by computing its time derivative.

\subsection{Non-linear non-Markovian dynamics}
\label{subsec:NLNMD}

Equation \eqref{eq:sse_linear} does not preserve the norm of the state vector $\ket{\phi_\wb}$. The second step of the construction thus consists in a subtle and often misapplied normalization. The normalized state vector
\begin{equation}\label{eq:normalization_state}
    \ket{\psi_\wb(t)} = \frac{1}{\sqrt{\langle \phi_\wb(t) |\phi_\wb(t)\rangle}}\ket{\phi_\wb(t)}
\end{equation}
no longer has an associated linear master equation \eqref{eq:master}, \ie~$\widetilde{\rho}(t) = \mathds{E}\big[\ket{\psi_\wb(t)}\bra{\psi_\wb(t)}\big]$ now has a non-trivial non-linear evolution. One approach is to define a physical or ``cooked'' \cite{bassi2003} probability measure $\upd \mathds{P}(t) =  \langle \phi_\wb(t) |\phi_\wb(t)\rangle \, \upd \mathds{Q} $ such that: 
\begin{equation}\label{eq:change_measure}
\rho(t) = \mathds{E}_\mathds{Q}\left[ \, \ket{\phi_\wb(t)}\bra{\phi_\wb(t)} \, \right] = \mathds{E}_{\mathds{P}(t)}\left[\,\ket{\psi_\wb(t)}\bra{\psi_\wb(t)}\,\right].
\end{equation}
This change of measure, proposed in the collapse literature \cite{bassi2002,bassi2003,adler2007,adler2008,ferialdi2012}, works \emph{a priori} only for a single time $t$, and is not sufficient to derive a consistent stochastic trajectory for states\footnote{The confusion comes from the fact that in the Markovian case, one has 
\begin{equation}\label{eq:markovian_consistency}
\forall T\geq t \; \mathds{E}_{\mathds{P}(T)}\left[\,\ket{\psi_\wb(t)}\bra{\psi_\wb(t)}\,\right] = \mathds{E}_{\mathds{P}(t)}\left[\,\ket{\psi_\wb(t)}\bra{\psi_\wb(t)}\,\right].
\end{equation} 
Hence, one can define the physical probability measure as $\mathds{P}(+\infty)$ and the condition \eqref{eq:change_measure} will stay valid for all time. In the non-Markovian case, \eqref{eq:markovian_consistency} is no longer true, physical probabilities $\mathds{P}(t)$ for different times disagree even for marginals restricted to their common pasts.}. 
The solution, well known in the open quantum system literature \cite{diosi1998,gambetta2002}, is instead to dynamically redefine a \emph{complete} noise trajectory $u\rightarrow \wb^{[t]}(u)$ at each $t$ such that for all functionals $f$:
\begin{equation}\label{eq:change_variable}
\mathds{E}_\mathds{Q}\left[f(\wb^{[t]}) \right] = \mathds{E}_{\mathds{P}(t)}\left[f(\wb) \right].
\end{equation}
The normalized quantum state trajectory $t \rightarrow \ket{\psi_{\wb^{[t]}}(t)}$ is then taken as the output of the model. It still has the same density matrix $\rho(t)$ upon averaging as the linearly evolving state:
\begin{equation}
    \rho(t) =\mathds{E}_\mathds{Q}\left[ \, \ket{\phi_\wb(t)}\bra{\phi_\wb(t)} \, \right] = \mathds{E}_\mathds{Q}\left[ \, \ket{\psi_{\wb^{[t]}}(t)}\bra{\psi_{\wb^{[t]}}(t)} \, \right]
\end{equation}
The final step is to find an explicit expression for $\wb^{[t]}$ as a function of $\wb$ using the relation \eqref{eq:change_variable}. After a rather non-trivial derivation shown in the appendix, one obtains: 
\begin{equation}\label{eq:explicit_change}
    w^{[t]}_k(v) = w_k(v) + 2 \sqrt{\gamma} \int_0^t \upd u\,  D_{jk}(u, v) \langle \hA_j\rangle_u,
\end{equation}
where 
\begin{equation}
\langle \hA_j\rangle_u = \frac{\bra{\phi_{\wb^{[u]}}(u)} \hA_j \ket{\phi_{\wb^{[u]}}(u)}}{\langle \phi_{\wb^{[u]}}(u)|\phi_{\wb^{[u]}}(u)\rangle} = \bra{\psi_{\wb^{[u]}}(u)} \hA_j \ket{\psi_{\wb^{[u]}}(u)}.
\end{equation}
Note that each $w_k^{[t]}(v)$ for $v\leq t$ does depend on states at times $u$ such that $v\leq u \leq t$. However, constructing a complete trajectory $w_k^{[t]}$ fortunately requires knowing \emph{only} the past states $\ket{\psi_{\wb^{[s]}}(s)}, \, s< t$, not future ones. 

 To summarize, the  
 non-linear stochastic trajectory $t\rightarrow \ket{\psi_{\wb^{[t]}}(t)}$  is defined by 
 the linear stochastic Schr\"odinger equation \eqref{eq:sse_linear}, the normalization \eqref{eq:normalization_state} and the continuous change of noise field \eqref{eq:explicit_change}. 
  Note, however, that this is an implicit definition. In particular, the trajectory is not straightforwardly `driven' by the noise process 
 $t\rightarrow \wb^{[t]}(t)$. 
 Rather, 
 to compute a trajectory $t\rightarrow \ket{\psi_{\wb^{[t]}}(t)}$,  one needs to know the noise $\wb^{[t]}(s)$, for all $s\leq t$ and \emph{not} only $\wb^{[s]}(s)$. From equation \eqref{eq:explicit_change}, one sees that to compute $\wb^{[t]}(s)$, one further needs to know all $\wb^{[t']}(s')$ for  $s'\leq t'\leq t$. Hence for every 
$\upd t$, one needs to compute a new \emph{complete} noise trajectory $ s\to  \wb^{[t]}(s)$ using \eqref{eq:explicit_change}, and construct the associated  (unnormalized) state $\ket{\phi_{\wb^{[t]}}(t)}$  using \eqref{eq:sse_linear}. This complicated procedure is explained in detail, and illustrated with a numerical example, in \cite{tilloy2017}.

All of the above complication  disappears in the Markovian limit where $D_{ij}(t,s)\rightarrow d_{ij} \delta(t-s)$. In that case $\wb^{[t]}(s) = \wb^{[s]}(s)$ for all $s\leq t$ and $\ket{\psi_{\wb^{[t]}}(t)}$ is really a simple  functional of $\wb^{[s]}(s)$ for 
$s \leq t$.  Further assuming that $d_{ij}=\delta_{ij}$, equations \eqref{eq:sse_linear} and \eqref{eq:explicit_change}  yield the It\^o equation \eqref{eq:sse_markov} for the normalized state (\ref{eq:normalization_state}),  where $\upd W_i(t) = w_i(t)  \,\upd t$. 

The dynamics in the non-Markovian setting are thus much more subtle than in the Markovian case, and far from simply driving the dynamics with colored noise. The steps presented above to find the normalized stochastic wave-function non-perturbatively are  non-trivial,
even for those familiar with linear and nonlinear stochastic Schr\"odinger equations  in the Markovian case. 
As we shall show in Sec.~\ref{sec:equiv}, the mapping to the Bohmian theory provides a considerably   easier way to understand it. 

\section{Bohmian mechanics}
Bohmian mechanics \cite{bohm1952I,bohm1952II,goldstein2016,durr2009} has \textit{a priori} nothing to do with collapse models, be they Markovian or non-Markovian. Consider a quantum system made of $N$ `particles' living in $d=1$ spatial dimension for simplicity, \ie~the Hilbert space is that of wave-functions $\psi(\xb)$ with $\xb\in\mathds{R}^N$. For such a system, Bohmian mechanics posits the existence of $N$ real particles with definite positions $\xb=\{x_k\}_{k=1\cdots N}$, whose velocities are given by a guiding equation:
 \begin{equation}
     \frac{\upd }{\upd t} x_k(t) = v_k(\xb(t),\ket{\psi_t}).
 \end{equation}
The velocity field $v_k$ is chosen in such a way that the Born rule is valid at every time if it is valid at the initial time (equivariance). More precisely, we now write $\hat{\mathbf{X}}=\{\hat{X}_k\}_{k=1}^N$ and $\hat{\mathbf{P}}=\{\hat{P}_k\}_{k=1}^N$ the associated position and momentum operators and consider spinless particles for simplicity. For a possibly time-dependent Hamiltonian $H(\hat{\mathbf{P}},\hat{\mathbf{X}})$ at most quadratic in the momenta, the canonical solution for the velocity field is to take \cite{gambetta2004}:
\begin{equation}\label{eq:velocityfield}
    v_k(\xb(t),\ket{\psi_t}) = \frac{\re\left[\langle\psi_t|\xb(t)\rangle\bra{\xb(t)} \hat{\mathcal{V}}_k\ket{\psi_t}\right]} {\langle\psi_t|\xb(t)\rangle\langle \xb(t)|\psi_t \rangle }
\end{equation}
where the velocity operator $\hat{\mathcal{V}}_k$ reads:
\begin{equation}\label{eq:velocityoperator}
    \hat{\mathcal{V}}_k= - i \left[\hat{X}_k, \hat{H}\right].
\end{equation}
If $\hat{H}$ is of the form:
\begin{equation}
    \hat{H} = \sum_{k=1}^N \frac{\hat{P}_k^2}{2 m_k} + V(\hat{\mathbf{X}}),
\end{equation}
then $\hat{\mathcal{V}}_k = \hat{P}_k/m_k$ and the prescription \eqref{eq:velocityfield} reduces to the more familiar Bohmian formula:
\begin{equation}
    v_k(\xb(t),\ket{\psi_t}) =\frac{1}{m_k} \im \left[\frac{ \partial_{x_k}\psi_t(\xb)}{\psi_t(\xb)} \right]\Big|_{\xb=\xb(t)},
\end{equation}
where $\psi_t(\xb) = \bra{\xb}\psi_t\rangle$.
Later on, we will need a mild extension of the previous general setup to a situation where there is an additional global degree of freedom to which no Bohmian variables are associated\footnote{Such non-Bohmian degrees of freedom have always been part of Bohmian mechanics; this is the standard treatment for spin~\cite{bohm1952I,bohm1952II}. Bolder precedents for this approach are discussed in Sec.~\ref{Sec:CMiSM}. The Hilbert space for these extra degree of freedom we consider is global (like those bolder precedents), rather than a tensor product of Hilbert spaces for each particle (as in the spin case).}, \ie~that the full Hilbert space is $\mathscr{H}_\text{full}=\mathscr{H}_\text{particles} \otimes \mathscr{H}_\text{aux}$. Writing $\hat{H}_\text{full}$ for the Hamiltonian of the full system including the extra degree of freedom, we have the same velocity \eqref{eq:velocityfield} as before:
\begin{align}
\hat{\mathcal{V}}_k&= - i \left[\hat{X}_k,  \hat{H}_\text{full}\right], \label{genVop} \\
    v_k(\xb(t),\ket{\psi_t}) &= \frac{\re\left[\langle\psi_t|\hat{\Pi}_\xb(t)  \hat{\mathcal{V}}_k\ket{\psi_t}\right]} {\langle\psi_t|\hat{\Pi}_\xb(t)|\psi_t \rangle },
    \label{genVval} 
\end{align}
where $\hat{\Pi}_\xb(t) = \ket{\xb(t)}\bra{\xb(t)} \otimes \mathds{1}_\text{aux}$ is the projector on particle positions. Now, let us emphasize two important points. First, in what follows, to derive the equivalence, the auxiliary Hilbert space $\mathscr{H}_\text{aux}$ that here seems idle, is going to be the system Hilbert space $\mathscr{H}_\text{sys}$ of the collapse model, while the Hilbert space containing the particles [here $L^2(\mathds{R}^N)$] is going to be that of the bath. Second, Eqs.~(\ref{genVop}) and (\ref{genVval}) are valid Bohmian dynamics for any hidden variable $\xb$ provided the Hamiltonian is at most quadratic in operators canonically conjugate to $\hat\xb$. This allows the Bohmian formalism to be applied to any linear combination of position and momenta for a bath of harmonic oscillators linearly coupled to an arbitrary system operator, as naturally arises. We will call any hidden variables with these dynamics Bohmian variables.

\section{Equivalence} \label{sec:equiv}

\subsection{Setup}
Our starting point is a generic collapse model with colored noise, whose dynamics are fully determined by equations \eqref{eq:sse_linear} and \eqref{eq:explicit_change}. As advertised, our objective is to construct a bath such that the dynamics of the system $+$ bath setup, where the bath includes Bohmian variables, is exactly that of the collapse model.

Let us consider a bath made of a continuum of harmonic oscillators parameterized by their frequency $\omega$ and an extra index $k$. The bath Hamiltonian\footnote{Note that this Hamiltonian is, strictly speaking, not bounded from below. It could easily be obtained as a limit of a more physical Hamiltonian, but that would complicate  
our derivation.} $\Hbath$ reads:
\begin{align}
\Hbath&=  \sum_{k=1}^D \int_\mathds{R} \upd \omega \,\omega\, a_{k,\omega}^\dagger\, a_{k,\omega}.
\end{align}
where the $a_{k,\omega}^\dagger$ and $a_{k,\omega}$ are the standard creation and annihilation operators $[a_{k,\omega},a_{k',\omega'}] = \delta_{kk'} \delta(\omega-\omega')$.
A quantum system of interest, of proper Hamiltonian $\Hsys$, is linearly coupled with the bath by the interaction Hamiltonian:
\begin{equation}
    \Hint =  \sqrt{\gamma} \,  \sum_{k=1}^D\, \hA_k \otimes \int_\mathds{R} \upd \omega \,\kappa^\ell_{k,\omega}\, \hp_{\ell,\omega} 
\end{equation}
where $\kappa^\ell_{k,\omega} = \kappa^{\ell}_{k,-\omega}$ are arbitrary real coefficients and we have introduced the reduced position and momentum operators:
\begin{align}
    \hx_{\ell,\omega} &= \frac{a_{\ell,\omega} + a^\dagger_{\ell,\omega}}{\sqrt{2}}\\
    \hp_{\ell,\omega} &=\frac{a_{\ell,\omega} - a^\dagger_{\ell,\omega}}{i\sqrt{2}}
\end{align}
We now write the full state  system + bath in the interaction picture: $\ket{\Psi^{I}(t)} = \e^{i t (\Hsys+ \Hbath)}\ket{\Psi(t)}$. For an initial state of the form $\ket{\psi(0)}\otimes \ket{0}$ where $\ket{0}$ is the bath state with zero excitations (Fock vacuum), one can integrate out the bosonic degrees of freedom using Wick's theorem (see \eg~\cite{diosi2014}):
\begin{align}
    \rho^{I}(t) &= \tr_{\text{bath}} \left[ \mathcal{T} \exp \left(-i\int_0^t \upd \tau\, \left[ \overrightarrow{H}_\text{int}(\tau)-\overleftarrow{H}_\text{int}(\tau)\right]\right)\ket{\Psi^{I}(0)} \bra{\Psi^{I}(0)}\right] \\
    &=\mathcal{M}_t\cdot \rho^{I}(0),
\end{align}
where the completely positive map $\mathcal{M}_t$ is given by \eqref{eq:master} provided we take $\kappa$ such that
\begin{equation}\label{eq:D}
    D_{jk}(t,s) = \sum_{\ell=1}^d\int_{\mathds{R}^+} \upd \omega\, \kappa^\ell_{j,\omega} \kappa^\ell_{k,\omega} \cos(\omega(t-s))\, 
\end{equation}
which is always possible if $D$ is time-translation invariant as we have assumed.
Hence, averaging over the noise in the collapse model and tracing out the bath in the system $+$ bath setup yields the same master equation. This is still far from proving the deeper equivalence we advertised, but it establishes the bath model we need.

\subsection{Equivalence at the linear equation level}\label{subsec:lineq}
Let us now discuss the Bohmian trajectories associated with the bath harmonic oscillators. The particles variables we shall consider are not directly the ones associated with the position operators $\hx_{k,\omega}$.  
Conceptually, there is no problem with considering the system dynamics conditioned on this, most obvious, choice of Bohmian variables. The average evolution of the conditioned wavefunction will be the same regardless. However, the dynamics of the individual Bohmian-conditioned system states is quite unlike that of the usual ``collapse'' evolution. In particular, even in the limit where the average dynamics is Markovian, the Bohmian-conditioned wavefunction evolution is not that of a Markovian (white noise) collapse model. For more discussion, see Refs.~\cite{GamWis03a,GamWis03}.

To obtain bath variables suitable for our purpose, we first consider the evolution in the interaction representation for the bath only. This is equivalent to considering the usual Bohmian positions for the time dependent Hamiltonians $\Hbath'=0$, and $\Hint'(t)= \e^{i t\Hbath} \Hint \e^{-i t \Hbath}$. Second we shall define new canonical pairs of variables:
\begin{align}
    \hx^{+}_{j,\omega} = \frac{\hx_{j,\omega} + \hx_{j,-\omega}}{\sqrt{2}} &\hskip1cm \hp^{+}_{j,\omega} = \frac{\hp_{j,\omega} + \hp_{j,-\omega}}{\sqrt{2}} \\
    \hx^{-}_{j,\omega} = \frac{\hp_{j,\omega} - \hp_{j,-\omega}}{\sqrt{2}} &\hskip1cm \hp^{-}_{j,\omega} = \frac{-\hx_{j,\omega} + \hx_{j,-\omega}}{\sqrt{2}}
\end{align}
These new operators indeed have the commutation relations:
\begin{align}
    [\hx^{+}_{j,\omega}, \hp^{+}_{j,\omega}] = i & \hskip1cm [\hx^{+}_{j,\omega}, \hp^{-}_{j,\omega}]=0\\
    [\hx^{-}_{j,\omega}, \hp^{+}_{j,\omega}] = 0 & \hskip1cm [\hx^{-}_{j,\omega}, \hp^{-}_{j,\omega}]=i,
\end{align}
and $x^{+}$ and $x^{-}$ can be formally regarded as position operators. 
The initial Hamiltonian can be rewritten as a function of these new canonical pairs, and the new oscillators explicitly interact between each other in this basis.
The new variables $(x^+,x^-)$, associated to the operators $\hx^+$ and $\hx^-$ are the one to which we associate Bohmian trajectories $x^+(t),x^-(t)$. We denote the collection of all these, for all the $k$ and all $\omega >0$, by $\xb(t)$.

Let us first consider the unnormalized system wave-function conditioned on arbitrary \emph{fixed} bath hidden variables:
\begin{equation}
    \ket{\widetilde{\phi}_\xb(t)} = \bra{\xb } \Phi^I(t) \rangle.
\end{equation}
We will need a variation of this conditional wave function that is normalized at the initial time:
\begin{align}
    \ket{\phi_\xb(t)} &=\frac{1}{ \bra{\xb} 0 \rangle} \bra{\xb} \Phi^I(t) \rangle \\
    &= \exp\left( \sum_{j=1}^d\int_{\mathds{R}^+} \upd \omega\, \frac{x^{+ 2}_{j,\omega}+x^{- 2}_{j,\omega}}{2}\right) \ket{\widetilde{\phi}_\xb(t)}.\label{eq:weirdnormalization} 
\end{align}
Let us compute the dynamics of this conditional wave-function, for fixed bath hidden variables.
The partial time derivative is simply:
\begin{align}
    \partial_t \ket{\widetilde{\phi}_{\xb}(t)} &= \bra{\xb} \partial_t\Phi^I(t) \rangle\\
    &=- i \Hsys \ket{\widetilde{\phi}_{\xb}(t)} -i\sqrt{\gamma} \int_{\mathds{R}}\upd\omega\, \kappa^\ell_{k,\omega} \, \hA_k \,\bra{\xb} \hp_{\ell,\omega}(t) \ket{\Phi^I(t)}  \\
    &=- i \Hsys \ket{\widetilde{\phi}_{\xb}(t)} -i\sqrt{\gamma} \int_{\mathds{R}}\upd\omega\, \kappa^\ell_{k,\omega} \, \hA_k \,\bra{\xb} \cos(\omega t) \hp_{\ell,\omega}(0) -\sin(\omega t) \hx_{\ell,\omega}(0) \ket{\Phi^I(t)}\, .
    \end{align}
    Now, splitting $\int_\mathbb{R}$ into $\int_{\mathbb{R}^+}+\int_{\mathbb{R}^-}$, changing of variable $\tilde{\omega}=-\omega$ in the second integral, regrouping the integrals using $\kappa^\ell_{k,\omega}=\kappa^\ell_{k,-\omega}$ and finally expressing the old canonical variables as a function of the new gives
    \begin{align}
   \partial_t \ket{\widetilde{\phi}_{\xb}(t)} &=- i \Hsys \ket{\widetilde{\phi}_{\xb}(t)} -i\sqrt{2\gamma} \int_{\mathds{R}^+}\upd\omega\, \kappa^\ell_{k,\omega} \, \hA_k \,\bra{\xb} \cos(\omega t) \hp^+_{\ell,\omega}(0) +\sin(\omega t) \hp^-_{\ell,\omega}(0) \ket{\Phi^I(t)} \\
    &= \bigg[- i \Hsys +\sqrt{2\gamma}\, \hA_k \int_{\mathds{R}^+}\!\!\! \upd \omega \, \kappa^\ell_{k,\omega}\Big(
    - \cos(\omega t)\frac{\delta}{\delta x^+_{\ell,\omega}}-\sin(\omega t)\frac{\delta}{\delta x^-_{\ell,\omega}}\Big)\bigg]  \ket{\widetilde{\phi}_{\xb}(t)}
\end{align}
where we sum on the repeated indices $\ell,k$ everywhere.
Using \eqref{eq:weirdnormalization}, we get
\begin{equation}\label{eq:linearbohm}
\begin{split}
    \partial_t \ket{\phi_{\xb}(t)} = \bigg[- i \Hsys +\sqrt{2\gamma}\, \hA_k \int_{\mathds{R}^+}\!\!\! \upd \omega \, \kappa^\ell_{k,\omega}&\Big(\cos(\omega t) x^+_{\ell,\omega}+\sin(\omega t)x^-_{\ell,\omega}\\
    &- \cos(\omega t)\frac{\delta}{\delta x^+_{\ell,\omega}}-\sin(\omega t)\frac{\delta}{\delta x^-_{\ell,\omega}}\Big)\bigg]  \ket{\phi_{\xb}(t)}.
    \end{split}
\end{equation}
We now define a field $w_i(\xb,s)$ obtained from the bath hidden variables, which as we will later see can be made equal to the noise of the collapse model:
\begin{equation}\label{eq:linknoiseposition}
    w_k(\xb,s) = \sqrt{2} \int_{\mathds{R}^+} \!\!\upd \omega\, \kappa^\ell_{k,\omega} \left(\cos(\omega s) x^+_{\ell,\omega}+\sin(\omega s)x^-_{\ell,\omega} \right). 
\end{equation}
Let us now rewrite \eqref{eq:linearbohm} as a function this field instead of the bath variables $\xb$. The functional derivatives of the field with respect to bath variables are
\begin{align}
    \frac{\delta w_k(\xb,s)}{\delta x_{j,\omega}^+} &=\sqrt{2}\kappa^j_{k,\omega} \cos(\omega s),\\
    \frac{\delta w_k(\xb,s)}{\delta x_{j,\omega}^-} &= \sqrt{2}\kappa^j_{k,\omega} \sin(\omega s).
\end{align}
Using the chain rule for functional derivatives and recalling the expression for $D$ in \eqref{eq:D}
we get
\begin{align}
\partial_t \ket{\phi_{\xb}(t)} &= \left[-i \hH  + \sqrt{\gamma} \, w_j(t) \hA_j  - 2 \sqrt{\gamma}\, \hA_j  \int_\mathds{R}  \upd s \, D_{jk}(t,s) \frac{\delta}{\delta w_k(s)} \right] \ket{\phi_{\xb}(t)}\\
&= \left[-i \hH  + \sqrt{\gamma} \, w_j(t) \hA_j  - 2 \sqrt{\gamma}\, \hA_j  \int_0^t  \upd s \, D_{jk}(t,s) \frac{\delta}{\delta w_k(s)} \right] \ket{\phi_{\xb}(t)},
\end{align}
where in the last line, we just change the bounds of the integral because $\frac{\delta}{\delta w_k(s)}\ket{\phi_{\xb}(t)}=0$ if $s>t$ (the noise $w_j(t)$ appears in the dynamics only at time $t$, and thus the state at $t$ does not depend explicitly on $w_j(s),\, s>t$). Hence, the unnormalized Bohmian wave function conditioned on \emph{fixed} bath hidden variables  $\ket{\phi_{\xb}(t)}$ is equal to the linear stochastic wave function of the collapse model $\ket{\phi_\wb(t)}$, where the link between $\xb$ and the noise field $\wb$ is given by equation \eqref{eq:linknoiseposition}.  

\subsection{Initial conditions and law of the field}
Let us now verify that the field $w_k(\xb,s)$ we have defined in \eqref{eq:linknoiseposition} has the same probability distribution as the noise driving the linear equation of the collapse model we are interested in.

To this end, we need to assume that the bath Bohmian variables are sampled according to the Born rule. Because we start in the ground state of all harmonic oscillators, the wave-function is Gaussian and hence the hidden variables are Gaussian random variables of zero mean, with 
\begin{align}
    \mathds{Q}(\xb) &=  |\bra{\xb} 0 \rangle|^2 \\
    &=\mathcal{N}^{-1} \exp\left(- \sum_{j=1}^d\int_{\mathds{R}^+} \upd \omega\, x^{+ 2}_{j,\omega}+x^{- 2}_{j,\omega}\right).\label{eq:bohmproba}
\end{align}
The noise field $w_k(\xb,s)$ is a linear functional of these random variables, and hence is also Gaussian of zero mean. Let us compute its two-point correlation function:
\begin{equation}
\begin{split}
    \mathds{E}[w_j(\xb,u) w_k(\xb,v)] = 2 \sum_{\ell_1,\ell_2 = 1}^d\int_{\mathds{R}^{+ 2}} \!\!\!\upd \omega_1 \upd \omega_2 \, \kappa_{j,\omega_1}^{\ell_1}\kappa_{k,\omega_2}^{\ell_2} &\bigg\{\cos (\omega_1 u) \cos( \omega_2 v) \mathds{E}[x^+_{k,\omega_1}
    x^+_{j,\omega_2}]\\
    &+ \sin (\omega_1 u) \sin (\omega_2 v) \mathds{E}[x^-_{j,\omega_1} x^-_{k,\omega_2}] \bigg\}.
    \end{split}
\end{equation}
From \eqref{eq:bohmproba} we get $\mathds{E}[x^+_{k,\omega_1}
    x^+_{j,\omega_2}]=\mathds{E}[x^-_{j,\omega_1} x^-_{k,\omega_2}] = \delta_{jk} \delta(\omega_1-\omega_2)$ and hence finally
\begin{equation}
    \mathds{E}_{\mathds{Q}}[w_j(\xb,u) w_k(\xb,v)] = D_{jk}(u,v).
\end{equation}
So, as advertised, the field we defined from Bohmian particle positions at the initial time has the same law as the collapse noise, provided the  Bohmian variables are statistically distributed according to the Born rule.

\subsection{Full equivalence}

So far, we have kept the Bohmian variables fixed, that is, we have considered $\ket{\widetilde{\psi}_\xb(t)}$ where $\xb=\xb(0)$ denotes the Bohmian variables associated to the bath oscillator at $t=0$. Let us now compute their velocity field, i.e. $v_{k,\omega}^{\pm} =\frac{\upd}{\upd t} x^\pm_{k,\omega}(t)$ as a function of the wave-function. Using  \eqref{eq:velocityoperator} we get
\begin{align}
    \hat{\mathcal{V}}^+_{k,\omega} &= \sqrt{2\gamma}\,  \kappa^k_{\ell,\omega} \cos (\omega t) \,  \hA_\ell \otimes \mathds{1},\\
    \hat{\mathcal{V}}^-_{k,\omega} &= \sqrt{2\gamma}\,  \kappa^k_{\ell,\omega} \sin (\omega t) \,  \hA_\ell \otimes \mathds{1},
\end{align}
and thus, using \eqref{eq:velocityfield},
\begin{align}
    v_{k,\omega}^+ &= \sqrt{2\gamma} \, \cos(\omega t)\, \kappa^k_{\ell,\omega} \, \frac{\bra{\phi_{\xb(t)}(t)} \hA_\ell \ket{\phi_{\xb(t)}(t)}}{\bra{\phi_{\xb(t)}(t)} \phi_{\xb(t)}(t)\rangle},\\
    v_{k,\omega}^- &= \sqrt{2\gamma}\,  \sin(\omega t)\, \kappa^k_{\ell,\omega}\,  \frac{\bra{\phi_{\xb(t)}(t)} \hA_\ell \ket{\phi_{\xb(t)}(t)}}{\bra{\phi_{\xb(t)}(t)} \psi_{\xb(t)}(t)\rangle}.
\end{align}
This allows us to compute $w_k(\xb(t),s)$, indeed
\begin{align}
    \frac{\upd}{\upd t} w_k(\xb(t),s) &=\sqrt{2} \int_{\mathds{R}^+} \!\!\upd \omega\, \kappa^\ell_{k,\omega} \left(\cos(\omega s) \frac{\upd x^+_{\ell,\omega}(t)}{\upd t}+\sin(\omega s) \frac{\upd x^-_{\ell,\omega}(t)}{\upd t} \right) \\
    &=2 \sqrt{\gamma} \int_{\mathds{R}^+} \upd \omega\, \kappa^\ell_{k,\omega} \kappa^\ell_{k',\omega} \,\cos (\omega(t-s))\, \langle \hA_{k'}\rangle_t\\
    &= 2\sqrt{\gamma} \, D_{k k'}(t,s)  \, \langle \hA_{k'}\rangle_t\, ,
\end{align}
where we have introduced the notation, similar to the one we used in the collapse case
\begin{equation}
    \langle \hat{A}_{k}\rangle_t := \frac{\bra{\phi_{\xb(t)}(t)}\hat{A}_k\ket{\phi_{\xb(t)}(t)}}{\langle \phi_{\xb(t)}(t)\ket{\phi_{\xb(t)}(t)}} = \bra{\psi_{\xb(t)}(t)}\hat{A}_k\ket{\psi_{\xb(t)}(t)}.
\end{equation}
Hence the time evolution of this function  $w_k(\xb(t),s)$ of bath hidden variables is exactly the same as that of the redefined noise field of the collapse model. Since for a given realization of noise (or fixed $\xb$), the Bohmian conditional wave function and the collapse model linear wave function are identical, we get that the full non-linear trajectories are identical, \ie 
\begin{equation}
\begin{split}
  \forall t \geq 0,\;  &\text{if} ~~~ w^{[0]}_k(u) = w_k(\xb(0),u) = \sqrt{2} \int\!\upd \omega\, \kappa^\ell_{k,\omega} \big(\cos(\omega u) x^+_{\ell,\omega}+\sin(\omega u)x^-_{\ell,\omega} \big)\\
  &\text{then} ~~~ \ket{\psi_{\wb^{[t]}}(t)}=\ket{\psi_{\xb(t)}(t)}
\end{split}
\end{equation}
This is the result we had advertised.

\subsection{Summary}
To summarize the equivalence we have derived, we provide a dictionary to go from one representation to the other. 

\vskip0.5cm

\begin{tabular}{Sl|Sl S{p{9cm}}}
     \textbf{Collapse model}  & \textbf{Bohmian reformulation} \\ \hline
  \makecell[l]{Gaussian noise field before redefinition  \\ $w_k(s)$ with $\mathds{E}_\mathds{Q}\left[w_j(t) w_k(s)\right] = D_{jk}(t,s)$}
  & \makecell[l]{Deterministic function of fixed bath  hidden variables \\     $w_k(\xb,s) = \sqrt{2} \int\!\upd \omega\, \kappa^\ell_{k,\omega} \big(\cos(\omega s) x^+_{\ell,\omega}+\sin(\omega s)x^-_{\ell,\omega} \big) $}\\ 
  \makecell[l]{Non-Gaussian noise field after redefinition \\ $    w^{[t]}_k(s) = w_k(s) + 2 \int_0^t\upd u\,  D_{k'k}(u, s) \langle A_{k'}\rangle_u,$}
  & \makecell[l]{Deterministic function of real-time bath hidden variables \\     $w_k(\xb(t),s) = \sqrt{2} \int \!\upd \omega\, \kappa^\ell_{k,\omega} \big(\cos(\omega s) x^+_{\ell,\omega}(t)+\sin(\omega s)x^-_{\ell,\omega}(t) \big) $}\\ 
  \makecell[l]{Intrinsic randomness progressively  \\introduced in the dynamics by noise} & \makecell[l]{Deterministic dynamics with uncertainty coming from \\ lack of knowledge of bath hidden variables at $t=0$}\\
  
  \makecell[l]{Un-normalized wave-function $\ket{\phi_\wb(t)}$ \\
  obeying the linear stochastic equation}& \makecell[l]{Wave-function conditioned on fixed bath particle positions \\ $\ket{\phi_\xb(t)} =\frac{1}{ \bra{\xb} 0 \rangle} \bra{\xb} \Phi^I(t) \rangle$} \\ 
  \makecell[l]{Normalized wave-function $\ket{\psi_{\wb^{[t]}} (t)}$ \\ obeying non-linear stochastic dynamics} & \makecell[l]{Normalized wave-function $\ket{\psi_{\xb(t)}(t)}$  \\ conditioned on real-time bath hidden variables }\\ 
\end{tabular}

\section{Discussion}

\subsection{Testability}
Conventionally, there has existed a distinction between strict {\em interpretations} of quantum mechanics, such as Bohmian mechanics, and {\em modifications}, such as collapse theories. 
The latter are supposed to be empirically distinguishable from standard quantum mechanics, a feature which is seen as desirable by many proponents. How, then, can this distinction make sense if, as we have shown, generic collapse theories are most easily understood as the conditional wave-function dynamics arising from a Bohmian type of theory? 

The answer to that question is that the class of 
collapse models we considered {\em are} consistent with quantum theory understood broadly because they are derived mathematically by coupling the physical system to another quantum mechanical system, a bath.
(In the special case of Markovian noise, this quantum bath can be thought of as a measurement apparatus). This is the only known way to obtain a collapse model that preserves the nice properties of quantum mechanics, such as the linearity of probability as a function of the density operator $\rho$, and the impossibility of faster-than-light signalling. That is, assuming that the collapse model accounts for the appearance of effectively classical measurement results when applied to macroscopic systems, its predictions for the statistics of the results of measurements made in this way on some (typically microscopic) system of interest $S$ can be obtained simply by tracing over the bath introduced to derive the collapse model for $S$.

Thus, by design, it is impossible to experimentally distinguish the signature of a non-Markovian collapse model from that of (purely quantum mechanical) extra degrees of freedom. But to be consistent with {\em some} quantum mechanical model does not make collapse models consistent with our best current physical theories. Most blatantly,  the extra degrees of freedom, 
could break symmetries we expect all fundamental quantum theories to obey, like Lorentz invariance.  This should have clear experimental signatures, to the extent that even the supposed guarantee of no-signalling offered by this style of theory could be voided on microscopic length scales. This is so whether we think of the bath as a purely mathematical tool for deriving non-Markovian collapse equations, or as physically existing degrees of freedom with associated Bohmian variables which define the collapse.

\subsection{Collapse models within the Standard Model?}
\label{Sec:CMiSM}
The considerations of the preceding subsection raise the question of whether it would be possible to formulate a collapse model respecting (at the empirical level) Lorentz symmetry, and other desirable symmetries. The obvious strategy would be to make the bosonic bath a free quantum field with dynamics respecting those symmetries, with a Hamiltonian coupling to the `physical' system which is linear in the bath field operators~ \cite{tilloy2017qft}.  But then one realizes that we already have a quantum field with these  properties: the electromagnetic radiation (EMR) field with the relativistic QED coupling. That is, one could consider the EMR field to be the `bath', and all other fundamental fields, bosonic and fermionic, to be the `system'. A commuting set of half the degrees of freedom of the EMR field is assigned a real value at all points in space-time, by the usual Bohmian theory for bosonic quantum fields \cite{struyve2007_bosonic}. It is only the wave-function of the `system' that has non-Markovian stochastic dynamics, conditioned on the EMR hidden variables. This approach to defining a fundamental ontology has in fact been proposed~\cite{struyve2007_bosonic} as a way to avoid the difficulties with defining Bohmian trajectories for relativistic fermions. 

Now, the theory just described is not precisely of the form set out in the body of the paper. First, the Bohmian degrees of freedom of the EMR field are those for the vector potential (\eg\ its value in each $j,\omega$-mode), rather than quadrature variables mixing modes of different frequencies $x_{j,\omega}^\pm$ as we have considered. Second, the EMR operator that appears in the coupling Hamiltonian to charged matter is the vector potential itself, rather than conjugate momenta $\hat p_{\ell,\omega}$ to the positions which are assigned hidden variables. Third, the matter operators that appear in that coupling Hamiltonian are (mechanical) momentum or velocity operators, which are not strictly localized in space as the collapse operators $\hat A_k$ were assumed to be. Nevertheless, it is very plausible that the QED coupling to the EMR does lead to decoherence of macroscopic objects in a quasiclassical basis \cite{schlosshauer2019}, and that the information in the EMR about the matter configuration would be carried in the values of the vector potential. Thus there is enough similarity in structure that it is worth thinking about whether this particular Bohmian theory could be considered a collapse model for the state of (non-EMR) matter. This then raises the question of whether generalisations beyond EMR, to other gauge fields, could also be so considered.

\subsection{Physicality, the Markovian Limit, and Terminology}

As we have discussed, non-Markovian collapse models cannot be regarded as a straightforward non-linear driving of the Schr\"odinger equation, with the only difference from the Markovian case being that the noise is colored. Non-Markovianity runs much deeper, and the whole evolution is non-trivial to obtain. This raises the question of how nature would implement such a peculiar evolution, where it seems that linearity at the master equation level and the Born rule emerge almost from a conspiracy.

The Bohmian rewriting, on the other hand, seems to provide a mechanical understanding of how such apparently complicated functional differential equation and field redefinition could emerge. In the Bohmian unitary rewriting, complexity comes from the fact that we should not expect the conditional wave-function to obey a simple closed  equation, since it describes only part of the universe. The universe as a whole still obeys simple dynamics, just as in Bohmian mechanics.

The construction we presented still works in the Markovian limit, where the noise field driving the collapse model becomes white. But in this limit, the advantage in terms of simplicity goes from the Bohmian description to the collapse one. Indeed, in the Markovian limit, all the subtleties in the definition of the collapse dynamics vanish, and the collapse model wave-function obeys a simple closed stochastic differential equation. The Bohmian rewriting, on the other hand, requires a continuously infinite number of particles. Their erratic motion, uncorrelated in time in this limit, effectively recreates the white noise. 

Based on these considerations, we think it could be preferable to restrict the name ``collapse models'' to models driven by white noise. This is the regime where one can think of the quantum state (on a Cauchy surface) as the state of the universe, since it is sufficient to generate all possible futures with the correct probabilities. Out of this regime, the state of the universe has to be more than the quantum state. 
Indeed, the natural description, as we have shown, is one where the quantum state is extended (to other degrees of freedom) and supplemented with hidden variables (for those degrees of freedom) satisfying Bohmian equations. 

\subsection{Randomness, Determinism,  Bell, and Local Friendliness}

The equivalence uncovered in this paper, between non-Markovian wave-function evolution models and certain Bohmian-type models, also shines new light onto issues of randomness and determinism.   
It has been argued previously by Gisin \cite{gisin2019} that the difference between deterministic and stochastic theories lies only in where one puts the randomness. In stochastic theories, randomness develops in time, definite outcomes emerging progressively. In deterministic theories (like classical or Bohmian mechanics), all the randomness is hidden in the initial conditions (which are either unknowable or only fixable to finite precision). For aesthetic and philosophical reasons, Gisin favors the first option. But our formulation of non-Markovian wave-function evolution models suggest that they are more naturally thought of as gradually revealing randomness in initial conditions. 

Moreover, this has implications for (non)locality, in Bell's theorem~\cite{Bel64} and a more recent theorem~\cite{Bong2020}. Here we are concerned with locality as defined in Refs.~\cite{Jar84,WisCav17}, with the approximate meaning of parameter independence~\cite{Shi84}. This is the sense for which the additional assumption of determinism is necessary\footnote{This is as opposed to a different sense, that Bell formalized as {\em local causality}~\cite{bell1976}, for which no such additional assumption is required.} to obtain Bell inequalities~\cite{Bel64}. 
Here we understand determinism to mean, very roughly, that measurement results are determined by the state of the universe on any Cauchy surface prior to the measurement, or at least on any leaf in a preferred foliation. For theories which take the wave-function of the universe as an element of reality, it is a property of the leaf itself, not localized anywhere in space-time. 

For a non-Markovian stochastic wave-function  model, if one rejects the Bohmian-like ontology we have proposed and takes the ontology to inhere in the wave-function, it would seem that one would obviously end up with a non-deterministic theory. In fact, things are not so straight-forward. 
Say the non-Markovian noise is band-limited, as will necessarily be the case if the bath used to define it 
has a Hamiltonian that is bounded below\footnote{This is because our construction uses pairs of positive and negative frequency modes; see Sec.~\ref{subsec:lineq}.}.
Then the wave-function evolution will be 
analytic; its value a finite time in the future can be predicted with arbitrary accuracy from finitely many derivatives. Thus if we take the wave-function and its time derivatives to be real, then knowing these on one Cauchy surface would allow one to predict the (supposedly stochastic) future of the wave-function, including the measurement records it generates. 

An implication of the above is that, unlike the Markovian case,  at least these types of  non-Markovian stochastic wave-function evolution models must be nonlocal. This follows by Bell's theorem; if the model is deterministic (or can be arbitrarily well approximated by a model that is deterministic) then it must be nonlocal. 
Markovian collapse theories, by contrast, are local because 
these collapses 
can be considered a measurement ``from outside the universe'', which is just the way that measurements are treated (``from outside the system'') in standard operational quantum mechanics. As operational quantum mechanics is local, in the sense of satisfying parameter independence~\cite{Jar84,Shi84,WisCav17}, so are Markovian collapse models.

Beyond Bell's theorem, the above conclusions have implications for the recently proven ``Local Friendliness'' (LF)  theorem~\cite{Bong2020}. This theorem combines Bell's theorem with Wigner's friend scenario, to replace Bell's 1964 assumption of determinism with one of the absolute reality of observations by a `friend'. This assumption, called ``Absoluteness of observed events'' in Ref.~\cite{Bong2020}, is applied even when that observation can be unitarily reversed. This reversal is necessary to violate a LF inequality (the analogue of a Bell inequality), and is arguably possible if, as is widely believed, universal quantum computing is possible~\cite{WisCavRie21}. But since `true' (Markovian) collapse models respect locality, while non-Markovian collapse models do not necessarily respect it,  different conclusions can be drawn. If a LF inequality can be violated in an experiment involving an intelligent quantum computer, true collapse models entail that the observations of this  intelligent party are not, or at least not always, real in an absolute sense. By contrast this negative conclusion does not follow for non-Markovian stochastic wave-function models and other hidden variable models; their ontology  may include physical correlates of observations by this intelligent party, but this will depend on the details of the model and of the physical  implementation of the quantum computation~\cite{WisCavRie21}.

\subsection{Is there an alternative to guiding laws?}
It is quite striking that two {\em a priori} completely different approaches to solving the measurement problem of quantum mechanics in a realist way --general collapse models and Bohmian-like theories-- turn out to be essentially the same thing formally.  Moreover this correspondence is not limited to theories with continuous evolution: jump-dynamics of  discrete hidden variables~\cite{Bel84,Sud87}) can be used to define non-Markovian stochastic wave-function  models with discontinuous collapses~\cite{GamAskWis04}. 
 This raises the question of whether this type of approach exhausts the possibilities for ``unromantic'' interpretations~\cite{bell1992}. That is, are there consistent laws for statistics of local beables that cannot be formulated using guiding laws for the velocity of beables directly derived from a quantum mechanical probability current (or, in the discrete case, 
transition rates derived from quantum mechanical 
transition probabilities)? 

One serious attempt in this direction was made by Kent \cite{kent2017} (see \cite{tilloy2017qft} for a proposal in the same spirit), and consists in fixing the beables all at once. Taking the wave-function for all time up to the end of the universe (or a time arbitrarily far in the future), one defines the beables for earlier times {\em a posteriori}. In such an approach, the dynamics of the beables is not obtained forward in time, knowing the past, as a Cauchy problem, but depends on the future. This does not come without difficulties, as was noted by Marsh and Butterfield \cite{marsh2018depictions,butterfield2019}, but likely deserves to be explored further.

\paragraph*{Acknowledgments}
AT was partly supported for this work by a postdoctoral fellowship from the Alexander von Humboldt foundation. HMW was partly supported for this work 
by the FQXi Grant ``Quantum and consciousness: paths to experiment, and implications
for interpretations''. We thank Peter Morgan for comments on the preprint. HMW acknowledges the Yuggera people, the traditional owners of the land at at Griffith University on which this work was undertaken.

\vskip1cm
\appendix

\section*{Appendix: change of noise variables}
Our objective is to compute the explicit expression of the transformed noise trajectory $\wb^{[t]}$, given in equation \eqref{eq:explicit_change}, starting from the implicit prescription \eqref{eq:change_variable}:
\begin{equation}
\mathds{E}_\mathds{Q}\left[f(\wb^{[t]}) \right] = \mathds{E}_{\mathds{P}(t)}\left[f(\wb) \right].
\end{equation}
The following derivation follows the same strategy as the original one in \cite{diosi1998} but matches the notations of the present paper and is meant to make it self contained.
Let us choose a functional $f$ of the form $f(\wb) = \norm^{-1} g(\wb)$ where $\norm = \langle \phi_{\wb}(t) |\phi_{\wb}(t)\rangle$. By definition, $\mathds{E}_{\mathds{P}(t)}[f(\wb)]=\mathds{E}_\mathds{\mathds{Q}}[f(\wb)\norm]$, hence:
\begin{equation}\label{eq:defineg}
    \mathds{E}_\mathds{Q}\left[f(\wb^{[t]})\right] = \mathds{E}_\mathds{Q}\left[g(\wb)\right].
\end{equation}
Consequently, the right hand side is time independent and:
\begin{align}
    0 &= \frac{\upd}{\upd t} \mathds{E}_\mathds{Q}\left[f(\wb^{[t]})\right] \\
    &= \mathds{E}_\mathds{Q} \left[g(\wb^{[t]}) \partial_t\normt^{-1} + \int_0^t \upd s\frac{\upd w_j^{[t]}(s)}{\upd t}\frac{\delta}{\delta w_j^{[t]}(s)} \left(g(\wb^{[t]}) \normt^{-1}\right)\right] \label{eq:secondline}
\end{align}
The first term can be computed using \eqref{eq:sse_linear}:
\begin{equation}\label{eq:exp_rhs}
\begin{split}
    \partial_t \normt^{-1} =& -2 \sqrt{\gamma}\langle \hA_j\rangle_t \, \normt^{-1} w_j^{[t]}(t)\\
    &+ 2\sqrt{\gamma} \normt^{-2}\int_0^t\!\! \upd s \, D_{jk}(t,s) \frac{\delta}{\delta w_k^{[t]}(s)}\left[ \langle \hA_j\rangle_t \normt\right] ,
\end{split}
\end{equation}
where we have used the notation:
\begin{equation}
\langle \hA_j\rangle_t = \frac{\bra{\phi_{\wb^{[t]}}(t)} \hA_j \ket{\phi_{\wb^{[t]}}(t)}}{\langle \phi_{\wb^{[t]}}(t)|\phi_{\wb^{[t]}}(t)\rangle} = \bra{\psi_{\wb^{[t]}}(t)} \hA_j \ket{\psi_{\wb^{[t]}}(t)}.
\end{equation}
Let us compute the first term in \eqref{eq:secondline}.
The Furutsu-Novikov formula \cite{novikov1965,adler2007} gives:
\begin{equation}
 \forall \,\mathcal{F}, \;\;\; \mathds{E}_\mathds{Q}\left[\mathcal{F}(\wb) w_j(t)\right] = \int_0^t \upd s \, D_{jk}(t,s) \mathds{E}_\mathds{Q}\left[\frac{\delta}{\delta w_k(s)}\mathcal{F}(w)\right].
\end{equation}
It is easily extended from $\wb$ to $\wb^{[t]}$ using \eqref{eq:defineg} and yields:
\begin{align}
    \mathds{E}_\mathds{Q}\left[g(\wb^{[t]})\langle \hA_j\rangle_t \, \normt^{-1} w_j^{[t]}(t)\right]
    &=\int_0^t \upd s \, D_{jk}(t,s) \mathds{E}_\mathds{Q}\left[\normt^{-1} \frac{\delta}{\delta w^{[t]}_k(s)} \left[g(\wb^{[t]}) \langle \hA_j\rangle_t\right] \right]
\end{align}
Inserting this formula in \eqref{eq:secondline} gives:
\begin{equation}
    \mathds{E}_\mathds{Q}\left[g(\wb^{[t]})\partial_t\normt^{-1} \right] = \mathds{E}_\mathds{Q}\left[-2\sqrt{\gamma}\int_0^t \!\!\upd s D_{jk}(t,s) \langle \hA_j\rangle_t  \frac{\delta}{\delta w_k^{[t]}(s)}\left[g(\wb^{[t]}) \normt^{-1}\right]\right],
\end{equation}
hence: 
\begin{equation}
    \mathds{E}_\mathds{Q} \left[\int_0^t \!\!\upd s \left(\frac{\upd w^{[t]}_k(s)}{\upd t}- 2 \sqrt{\gamma} D_{jk}(t,s) \langle \hA_j\rangle_t \right) \frac{\delta}{\delta w_k^{[t]}(s)}\left[g(\wb^{[t]}) \normt^{-1}\right]\right]= 0,
\end{equation}
This latter expression is true for all functionals $g$ thus:
\begin{equation}
    \frac{\upd w^{[t]}_j(s)}{\upd t}- 2 \sqrt{\gamma} D_{jk}(t,s) \langle \hA_j\rangle_t=0
\end{equation}
and finally:
\begin{equation}
    w^{[t]}_k(s) = w_k(s) + 2\sqrt{\gamma}\int_0^t\upd \tau\,  D_{jk}(\tau, s) \langle \hA_j\rangle_\tau.
\end{equation}

\bibliographystyle{unsrtnat}
\bibliography{main,QMCrefsPLUS}

\begin{thebibliography}{60}
\providecommand{\natexlab}[1]{#1}
\providecommand{\url}[1]{\texttt{#1}}
\expandafter\ifx\csname urlstyle\endcsname\relax
  \providecommand{\doi}[1]{doi: #1}\else
  \providecommand{\doi}{doi: \begingroup \urlstyle{rm}\Url}\fi

\bibitem[Bohm(1952{\natexlab{a}})]{bohm1952I}
David Bohm.
\newblock A suggested interpretation of the quantum theory in terms of "hidden"
  variables. i.
\newblock \emph{Phys. Rev.}, 85:\penalty0 166--179, Jan 1952{\natexlab{a}}.
\newblock \doi{10.1103/PhysRev.85.166}.

\bibitem[Bohm(1952{\natexlab{b}})]{bohm1952II}
David Bohm.
\newblock A suggested interpretation of the quantum theory in terms of "hidden"
  variables. ii.
\newblock \emph{Phys. Rev.}, 85:\penalty0 180--193, Jan 1952{\natexlab{b}}.
\newblock \doi{10.1103/PhysRev.85.180}.

\bibitem[D{\"u}rr and Teufel(2009)]{durr2009}
Detlef D{\"u}rr and Stefan Teufel.
\newblock \emph{Bohmian mechanics: the physics and mathematics of quantum
  theory}.
\newblock Springer Science \& Business Media, Berlin, Germany, 2009.

\bibitem[Goldstein(2016)]{goldstein2016}
Sheldon Goldstein.
\newblock Bohmian mechanics.
\newblock In Edward~N. Zalta, editor, \emph{The Stanford Encyclopedia of
  Philosophy}. Metaphysics Research Lab, Stanford University, fall 2016
  edition, 2016.
\newblock URL
  \url{https://plato.stanford.edu/archives/fall2016/entries/qm-bohm/}.

\bibitem[Ghirardi et~al.(1986)Ghirardi, Rimini, and Weber]{ghirardi1986}
G.~C. Ghirardi, A.~Rimini, and T.~Weber.
\newblock Unified dynamics for microscopic and macroscopic systems.
\newblock \emph{Phys. Rev. D}, 34:\penalty0 470--491, Jul 1986.
\newblock \doi{10.1103/PhysRevD.34.470}.

\bibitem[Bassi and Ghirardi(2003)]{bassi2003}
Angelo Bassi and GianCarlo Ghirardi.
\newblock Dynamical reduction models.
\newblock \emph{Phys. Rep.}, 379\penalty0 (5):\penalty0 257 -- 426, 2003.
\newblock ISSN 0370-1573.
\newblock \doi{10.1016/S0370-1573(03)00103-0}.

\bibitem[Bassi et~al.(2013{\natexlab{a}})Bassi, Lochan, Satin, Singh, and
  Ulbricht]{bassi2013review}
Angelo Bassi, Kinjalk Lochan, Seema Satin, Tejinder~P. Singh, and Hendrik
  Ulbricht.
\newblock Models of wave-function collapse, underlying theories, and
  experimental tests.
\newblock \emph{Rev. Mod. Phys.}, 85:\penalty0 471--527, Apr
  2013{\natexlab{a}}.
\newblock \doi{10.1103/RevModPhys.85.471}.

\bibitem[Bell(1992)]{bell1992}
J.~S. Bell.
\newblock Six possible worlds of quantum mechanics.
\newblock \emph{Foundations of Physics}, 22\penalty0 (10):\penalty0 1201--1215,
  Oct 1992.
\newblock ISSN 1572-9516.
\newblock \doi{10.1007/BF01889711}.

\bibitem[Bell(1976)]{bell1976}
J.~S. Bell.
\newblock The theory of local beables.
\newblock \emph{Epistemological Letters}, 9\penalty0 (11), 1976.
\newblock (Reproduced in Ref.~\cite{BellCollection}.).

\bibitem[Allori(2015)]{allori2015}
Valia Allori.
\newblock Primitive ontology in a nutshell.
\newblock \emph{International Journal of Quantum Foundations}, 1\penalty0
  (2):\penalty0 107--122, 2015.
\newblock URL \url{http://www.ijqf.org/archives/2394}.

\bibitem[Allori et~al.(2008)Allori, Goldstein, Tumulka, and
  Zangh\`i]{allori2008}
Valia Allori, Sheldon Goldstein, Roderich Tumulka, and Nino Zangh\`i.
\newblock On the common structure of bohmian mechanics and the
  ghirardi–rimini–weber theory dedicated to giancarlo ghirardi on the
  occasion of his 70th birthday.
\newblock \emph{Br. J. Philos. Sci.}, 59\penalty0 (3):\penalty0 353--389, 2008.
\newblock \doi{10.1093/bjps/axn012}.

\bibitem[Allori et~al.(2014)Allori, Goldstein, Tumulka, and
  Zangh{\`\i}]{allori2014}
Valia Allori, Sheldon Goldstein, Roderich Tumulka, and Nino Zangh{\`\i}.
\newblock Predictions and primitive ontology in quantum foundations: a study of
  examples.
\newblock \emph{Br. J. Philos. Sci.}, 65\penalty0 (2):\penalty0 323--352, 2014.
\newblock \doi{10.1093/bjps/axs048}.

\bibitem[Toros et~al.(2016)Toros, Donadi, and Bassi]{toros2016}
Marko Toros, Sandro Donadi, and Angelo Bassi.
\newblock Bohmian mechanics, collapse models and the emergence of classicality.
\newblock \emph{J. Phys. A: Math. Theor.}, 49\penalty0 (35):\penalty0 355302,
  2016.
\newblock \doi{10.1088/1751-8113/49/35/355302}.

\bibitem[Gambetta and Wiseman(2003{\natexlab{a}})]{gambetta2003}
Jay Gambetta and H.~M. Wiseman.
\newblock Interpretation of non-markovian stochastic schr\"odinger equations as
  a hidden-variable theory.
\newblock \emph{Phys. Rev. A}, 68:\penalty0 062104, Dec 2003{\natexlab{a}}.
\newblock \doi{10.1103/PhysRevA.68.062104}.

\bibitem[Bassi and Ghirardi(2002)]{bassi2002}
Angelo Bassi and GianCarlo Ghirardi.
\newblock Dynamical reduction models with general gaussian noises.
\newblock \emph{Phys. Rev. A}, 65:\penalty0 042114, Apr 2002.
\newblock \doi{10.1103/PhysRevA.65.042114}.

\bibitem[Adler and Bassi(2007)]{adler2007}
Stephen~L. Adler and Angelo Bassi.
\newblock Collapse models with non-white noises.
\newblock \emph{J. Phys. A: Math. Theor.}, 40\penalty0 (50):\penalty0 15083,
  2007.
\newblock \doi{10.1088/1751-8113/40/50/012}.

\bibitem[Adler and Bassi(2008)]{adler2008}
Stephen~L. Adler and Angelo Bassi.
\newblock Collapse models with non-white noises: Ii. particle-density coupled
  noises.
\newblock \emph{J. Phys. A: Math. Theor.}, 41\penalty0 (39):\penalty0 395308,
  2008.
\newblock \doi{10.1088/1751-8113/41/39/395308}.

\bibitem[Bassi and Ferialdi(2009{\natexlab{a}})]{bassi2009exactshort}
Angelo Bassi and Luca Ferialdi.
\newblock Non-markovian quantum trajectories: An exact result.
\newblock \emph{Phys. Rev. Lett.}, 103:\penalty0 050403, Jul
  2009{\natexlab{a}}.
\newblock \doi{10.1103/PhysRevLett.103.050403}.

\bibitem[Bassi and Ferialdi(2009{\natexlab{b}})]{bassi2009exactlong}
Angelo Bassi and Luca Ferialdi.
\newblock Non-markovian dynamics for a free quantum particle subject to
  spontaneous collapse in space: General solution and main properties.
\newblock \emph{Phys. Rev. A}, 80:\penalty0 012116, Jul 2009{\natexlab{b}}.
\newblock \doi{10.1103/PhysRevA.80.012116}.

\bibitem[Ferialdi and Bassi(2012{\natexlab{a}})]{ferialdi2012exact}
Luca Ferialdi and Angelo Bassi.
\newblock Exact solution for a non-markovian dissipative quantum dynamics.
\newblock \emph{Phys. Rev. Lett.}, 108:\penalty0 170404, Apr
  2012{\natexlab{a}}.
\newblock \doi{10.1103/PhysRevLett.108.170404}.

\bibitem[Di\'osi(1989)]{diosi1989}
L.~Di\'osi.
\newblock Models for universal reduction of macroscopic quantum fluctuations.
\newblock \emph{Phys. Rev. A}, 40:\penalty0 1165--1174, Aug 1989.
\newblock \doi{10.1103/PhysRevA.40.1165}.

\bibitem[Pearle(1989)]{pearle1989}
Philip Pearle.
\newblock Combining stochastic dynamical state-vector reduction with
  spontaneous localization.
\newblock \emph{Phys. Rev. A}, 39:\penalty0 2277--2289, Mar 1989.
\newblock \doi{10.1103/PhysRevA.39.2277}.

\bibitem[Ghirardi et~al.(1990)Ghirardi, Pearle, and Rimini]{ghirardi1990}
Gian~Carlo Ghirardi, Philip Pearle, and Alberto Rimini.
\newblock Markov processes in hilbert space and continuous spontaneous
  localization of systems of identical particles.
\newblock \emph{Phys. Rev. A}, 42:\penalty0 78--89, Jul 1990.
\newblock \doi{10.1103/PhysRevA.42.78}.

\bibitem[Gisin(1989)]{gisin1989}
Nicolas Gisin.
\newblock Stochastic quantum dynamics and relativity.
\newblock \emph{Helv. Phys. Acta}, 62\penalty0 (4):\penalty0 363--371, 1989.
\newblock \doi{10.5169/seals-116034}.

\bibitem[Gisin(1990)]{gisin1990}
Nicolas Gisin.
\newblock Weinberg's non-linear quantum mechanics and supraluminal
  communications.
\newblock \emph{Phys. Lett. A}, 143\penalty0 (1-2):\penalty0 1--2, 1990.
\newblock \doi{10.1016/0375-9601(90)90786-N}.

\bibitem[Polchinski(1991)]{polchinski1991}
Joseph Polchinski.
\newblock Weinberg's nonlinear quantum mechanics and the
  einstein-podolsky-rosen paradox.
\newblock \emph{Phys. Rev. Lett.}, 66:\penalty0 397--400, Jan 1991.
\newblock \doi{10.1103/PhysRevLett.66.397}.

\bibitem[Bassi and Hejazi(2015)]{bassi2015}
Angelo Bassi and Kasra Hejazi.
\newblock No-faster-than-light-signaling implies linear evolution. a
  re-derivation.
\newblock \emph{Eur. J. Phys.}, 36\penalty0 (5):\penalty0 055027, 2015.
\newblock \doi{10.1088/0143-0807/36/5/055027}.

\bibitem[Bassi et~al.(2013{\natexlab{b}})Bassi, D\"urr, and
  Hinrichs]{bassi2013}
Angelo Bassi, Detlef D\"urr, and G\"unter Hinrichs.
\newblock Uniqueness of the equation for quantum state vector collapse.
\newblock \emph{Phys. Rev. Lett.}, 111:\penalty0 210401, Nov
  2013{\natexlab{b}}.
\newblock \doi{10.1103/PhysRevLett.111.210401}.

\bibitem[Wiseman and Di\'osi(2001)]{wiseman2001}
Howard~M. Wiseman and Lajos Di\'osi.
\newblock Complete parameterization, and invariance, of diffusive quantum
  trajectories for markovian open systems.
\newblock \emph{Chem. Phys.}, 268\penalty0 (1):\penalty0 91 -- 104, 2001.
\newblock ISSN 0301-0104.
\newblock \doi{10.1016/S0301-0104(01)00296-8}.

\bibitem[Myrvold(2017)]{myrvold2017}
Wayne~C. Myrvold.
\newblock Relativistic markovian dynamical collapse theories must employ
  nonstandard degrees of freedom.
\newblock \emph{Phys. Rev. A}, 96:\penalty0 062116, Dec 2017.
\newblock \doi{10.1103/PhysRevA.96.062116}.

\bibitem[Jones et~al.(2021)Jones, Guaita, and Bassi]{jones2021}
C.~Jones, T.~Guaita, and A.~Bassi.
\newblock Impossibility of extending the ghirardi-rimini-weber model to
  relativistic particles.
\newblock \emph{Phys. Rev. A}, 103:\penalty0 042216, Apr 2021.
\newblock \doi{10.1103/PhysRevA.103.042216}.

\bibitem[Tilloy(2017)]{tilloy2017}
Antoine Tilloy.
\newblock Time-local unraveling of non-{M}arkovian stochastic
  {S}chr{\"{o}}dinger equations.
\newblock \emph{{Quantum}}, 1:\penalty0 29, September 2017.
\newblock ISSN 2521-327X.
\newblock \doi{10.22331/q-2017-09-19-29}.

\bibitem[Di\'osi and Ferialdi(2014)]{diosi2014}
L.~Di\'osi and L.~Ferialdi.
\newblock General non-markovian structure of gaussian master and stochastic
  schr\"odinger equations.
\newblock \emph{Phys. Rev. Lett.}, 113:\penalty0 200403, Nov 2014.
\newblock \doi{10.1103/PhysRevLett.113.200403}.

\bibitem[Feynman and Vernon(1963)]{feynman1963}
Richard~Phillips Feynman and Frank~Lee Vernon.
\newblock The theory of a general quantum system interacting with a linear
  dissipative system.
\newblock \emph{Ann. Phys. (N Y)}, 24:\penalty0 118--173, 1963.
\newblock \doi{10.1016/0003-4916(63)90068-X}.

\bibitem[Strunz(1996)]{strunz1996}
Walter~T. Strunz.
\newblock Linear quantum state diffusion for non-markovian open quantum
  systems.
\newblock \emph{Phys. Lett. A}, 224\penalty0 (1):\penalty0 25 -- 30, 1996.
\newblock ISSN 0375-9601.
\newblock \doi{10.1016/S0375-9601(96)00805-5}.

\bibitem[Di{\'o}si and Strunz(1997)]{diosi1997}
Lajos Di{\'o}si and Walter~T Strunz.
\newblock The non-markovian stochastic schr{\"o}dinger equation for open
  systems.
\newblock \emph{Phys. Lett. A}, 235\penalty0 (6):\penalty0 569--573, 1997.
\newblock \doi{10.1016/S0375-9601(97)00717-2}.

\bibitem[Ferialdi and Bassi(2012{\natexlab{b}})]{ferialdi2012}
Luca Ferialdi and Angelo Bassi.
\newblock Dissipative collapse models with nonwhite noises.
\newblock \emph{Phys. Rev. A}, 86:\penalty0 022108, Aug 2012{\natexlab{b}}.
\newblock \doi{10.1103/PhysRevA.86.022108}.

\bibitem[Di\'osi et~al.(1998)Di\'osi, Gisin, and Strunz]{diosi1998}
L.~Di\'osi, N.~Gisin, and W.~T. Strunz.
\newblock Non-markovian quantum state diffusion.
\newblock \emph{Phys. Rev. A}, 58:\penalty0 1699--1712, Sep 1998.
\newblock \doi{10.1103/PhysRevA.58.1699}.

\bibitem[Gambetta and Wiseman(2002)]{gambetta2002}
Jay Gambetta and H.~M. Wiseman.
\newblock Non-markovian stochastic schr\"odinger equations: Generalization to
  real-valued noise using quantum-measurement theory.
\newblock \emph{Phys. Rev. A}, 66:\penalty0 012108, Jul 2002.
\newblock \doi{10.1103/PhysRevA.66.012108}.

\bibitem[Gambetta and Wiseman(2004)]{gambetta2004}
Jay Gambetta and H.~M. Wiseman.
\newblock Modal dynamics for positive operator measures.
\newblock \emph{Found. Phys.}, 34\penalty0 (3):\penalty0 419--448, Mar 2004.
\newblock ISSN 1572-9516.
\newblock \doi{10.1023/B:FOOP.0000019622.81881.f8}.

\bibitem[Gambetta and Wiseman(2003{\natexlab{b}})]{GamWis03a}
Jay Gambetta and Howard Wiseman.
\newblock A non-markovian stochastic schrodinger equation developed from a
  hidden variable interpretation.
\newblock \emph{Proceedings of SPIE - The International Society for Optical
  Engineering}, 5111, 05 2003{\natexlab{b}}.
\newblock \doi{10.1117/12.496938}.

\bibitem[Gambetta and Wiseman(2003{\natexlab{c}})]{GamWis03}
J.~Gambetta and H.~M. Wiseman.
\newblock Interpretation of non-{Markov}ian stochastic {S}chr\"odinger
  equations as a hidden-variable theory.
\newblock \emph{Phys. Rev. A}, 68\penalty0 (6):\penalty0 062104, Dec
  2003{\natexlab{c}}.
\newblock \doi{10.1103/PhysRevA.68.062104}.

\bibitem[{Tilloy}(2017)]{tilloy2017qft}
A.~{Tilloy}.
\newblock {Interacting quantum field theories as relativistic statistical field
  theories of local beables}.
\newblock \emph{arXiv:1702.06325}, 2017.
\newblock URL \url{https://arxiv.org/abs/1702.06325}.

\bibitem[Struyve and Westman(2007)]{struyve2007_bosonic}
W~Struyve and H~Westman.
\newblock A minimalist pilot-wave model for quantum electrodynamics.
\newblock \emph{Proc. R. Soc. A.}, 463\penalty0 (2088):\penalty0 3115--3129,
  2007.
\newblock \doi{10.1098/rspa.2007.0144}.

\bibitem[Schlosshauer(2019)]{schlosshauer2019}
Maximilian Schlosshauer.
\newblock Quantum decoherence.
\newblock \emph{Physics Reports}, 831:\penalty0 1--57, 2019.
\newblock ISSN 0370-1573.
\newblock \doi{10.1016/j.physrep.2019.10.001}.

\bibitem[Gisin(2019)]{gisin2019}
Nicolas Gisin.
\newblock Indeterminism in physics, classical chaos and bohmian mechanics: Are
  real numbers really real?
\newblock \emph{Erkenntnis}, Oct 2019.
\newblock ISSN 1572-8420.
\newblock \doi{10.1007/s10670-019-00165-8}.

\bibitem[Bell(1964)]{Bel64}
J.~S. Bell.
\newblock On the {Einstein}-{Podolsy}-{Rosen} paradox.
\newblock \emph{Physics}, 1:\penalty0 195, 1964.
\newblock (Reproduced in Ref.~\cite{BellCollection}.).

\bibitem[Bong et~al.(2020)Bong, Utreras-Alarc{\'o}n, Ghafari, Liang, Tischler,
  Cavalcanti, Pryde, and Wiseman]{Bong2020}
Kok-Wei Bong, An{\'\i}bal Utreras-Alarc{\'o}n, Farzad Ghafari, Yeong-Cherng
  Liang, Nora Tischler, Eric~G. Cavalcanti, Geoff~J. Pryde, and Howard~M.
  Wiseman.
\newblock A strong no-go theorem on the wigner's friend paradox.
\newblock \emph{Nature Physics}, 2020.
\newblock \doi{10.1038/s41567-020-0990-x}.

\bibitem[Jarrett(1984)]{Jar84}
Jon Jarrett.
\newblock On the physical significance of the locality conditions in the {Bell}
  argument.
\newblock \emph{No\^us}, 18:\penalty0 569--89, 1984.
\newblock \doi{10.2307/2214878}.

\bibitem[Wiseman and Cavalcanti(2017)]{WisCav17}
H.~M. Wiseman and E.~G. Cavalcanti.
\newblock {\em Causarum Investigatio} and the two {B}ell's theorems of {J}ohn
  {B}ell.
\newblock In Reinhold Bertlmann and Anton Zeilinger, editors, \emph{Quantum
  [Un]speakables II: Half a Century of Bell's Theorem}, The Frontiers
  Collection, pages 119--142, Switzerland, 2017. Springer.

\bibitem[Shimony(1984)]{Shi84}
A.~Shimony.
\newblock Controllable and uncontrollable non-locality.
\newblock In Susumu Kamefuchi, editor, \emph{Foundations of Quantum Mechanics
  in the Light of New Technology}, pages 225--230, Tokyo, 1984. Physical
  Society of Japan.

\bibitem[Wiseman et~al.(2021)Wiseman, Cavalcanti, and Rieffel]{WisCavRie21}
H.~M. Wiseman, E.~G. Cavalcanti, and Eleanor~G. Rieffel.
\newblock A `thoughtful' local friendliness no-go theorem.
\newblock in preparation, 2021.

\bibitem[Bell(1984)]{Bel84}
J.~S. Bell.
\newblock Beables for quantum field theory.
\newblock Technical Report TH.4035/84, CERN, Geneva, 1984.
\newblock (Reproduced in Ref.~\cite{BellCollection}.).

\bibitem[Sudbery(1987)]{Sud87}
A~Sudbery.
\newblock Objective interpretations of quantum mechanics and the possibility of
  a deterministic limit.
\newblock \emph{J. Phys. A: Math. Gen.}, 20\penalty0 (7):\penalty0 1743, 1987.
\newblock \doi{10.1088/0305-4470/20/7/020}.

\bibitem[Gambetta et~al.(2004)Gambetta, Askerud, and Wiseman]{GamAskWis04}
J.~Gambetta, T.~Askerud, and H.~M. Wiseman.
\newblock Jumplike unravelings for non-{Markov}ian open quantum systems.
\newblock \emph{Phys. Rev. A}, 69\penalty0 (5):\penalty0 052104, May 2004.
\newblock \doi{10.1103/PhysRevA.69.052104}.

\bibitem[Kent(2017)]{kent2017}
Adrian Kent.
\newblock Quantum reality via late-time photodetection.
\newblock \emph{Phys. Rev. A}, 96\penalty0 (6), Dec 2017.
\newblock ISSN 2469-9934.
\newblock \doi{10.1103/physreva.96.062121}.

\bibitem[Marsh(2018)]{marsh2018depictions}
Brendan Marsh.
\newblock Depictions of quantum reality in {K}ent's interpretation of quantum
  theory, 2018.
\newblock URL \url{https://arxiv.org/abs/1811.08950}.

\bibitem[Butterfield and Marsh(2019)]{butterfield2019}
J~Butterfield and B~Marsh.
\newblock Non-locality and quasiclassical reality in kent’s formulation of
  relativistic quantum theory.
\newblock \emph{Journal of Physics: Conference Series}, 1275:\penalty0 012002,
  Sep 2019.
\newblock ISSN 1742-6596.
\newblock \doi{10.1088/1742-6596/1275/1/012002}.

\bibitem[Novikov(1965)]{novikov1965}
E.~A. Novikov.
\newblock Functionals and the random-force method in turbulence theory.
\newblock \emph{JETP}, 20:\penalty0 5, 1965.
\newblock URL \url{http://www.jetp.ac.ru/cgi-bin/dn/e_020_05_1290.pdf}.

\bibitem[Bell et~al.(2001)Bell, Gottfried, and Veltman]{BellCollection}
M.~Bell, K.~Gottfried, and M.~Veltman, editors.
\newblock \emph{John S. Bell on the Foundations of Quantum Mechanics}.
\newblock World Scientific, Singapore, 2001.

\end{thebibliography}

\end{document}